\documentclass[aps]{revtex4-2}

\usepackage{graphicx}
\usepackage{natbib}
\usepackage{color}
\usepackage{floatrow}
\usepackage{subfig}
\usepackage{booktabs}
\usepackage{amssymb}%
\usepackage{amsmath}
\usepackage{multirow}
\usepackage{pgfplots}
  \let\leq=\leqslant
\let\ge=\geqslant  \let\geq=\geqslant
\usepackage{amsbsy}
\allowdisplaybreaks

\begin{document}

	\title{Spectrogram analysis of surface elevation signals due to accelerating ships}
	
	\author{Ravindra Pethiyagoda$^1$}
	\author{Timothy J. Moroney$^1$}
	\author{Gregor J. Macfarlane$^2$}
	\author{Scott W. McCue$^1$}
	
	\email[]{scott.mccue@qut.edu.au}
			
	\affiliation{$^1$School of Mathematical Sciences, Queensland University of Technology, QLD 4001, Australia}
	\affiliation{$^2$Australian Maritime College, University of Tasmania, Launceston, TAS 7248, Australia}
	\date{\today}
	
	\begin{abstract}
		Spectrograms provide an efficient way to analyse surface elevation signals of ship waves taken from a sensor fixed at a single point in space. Recent work based on a simplified model for the ship's disturbance suggests that matching the spectrogram heat-map patterns to a so-called dispersion curve has the potential for estimating of properties of a steadily moving ship, such as the ship's speed and closest distance to the sensor.  Here we extend the theory behind the dispersion curve so that it can be applied to ships accelerating along arbitrary paths and demonstrate how acceleration affects the structure of the associated spectrograms.  Examples are provided for a simple model of a ship accelerating/decelerating in a straight line or travelling in a circle with constant angular speed. We highlight a problem with non-uniqueness of the dispersion curve when comparing ships moving along different paths. Finally, we validate the new dispersion curve against experimental results of ship models accelerating in a finite depth basin.  Our work will provide a basis for more comprehensive studies that extend the simplified model to take into account the shape of the hull in question.
	\end{abstract}

	\maketitle

\section{Introduction}
A difficult problem in water wave theory is to measure the surface elevation at a single point in a shipping channel or open water and use only that data to deduce various properties of ships that pass by~\cite{pethiyagoda17}.  The properties we have in mind include the ships' velocities and distances from the sensor, the energy in each ship's wake, and the shape of the hull of each ship.  The possible applications of this type of research are numerous.  For example, predicting the energy from ship wakes in a shipping channel using only data from a fixed sensor can provide valuable information about ongoing shoreline erosion or possible damage to moored vessels \cite{didenkulova13,torsvik15a,safak20}.  Other applications relate to remote monitoring of waterways to determine whether vessels are complying with regulatory or operating conditions or even monitoring of illegal fishing boats or other unauthorised entry vessels.

From a mathematical perspective, this problem requires a time-frequency analysis of the given wave signal with a view to decoding the frequency spectrum in terms of recognisable contributions.  The tool we employ to achieve this goal is the spectrogram, which uses short-time Fourier transforms to decompose each wave signal from the fixed sensor into a time-frequency heat-map.  In recent times, some success has been achieved in using spectrograms to identify different features of ship wakes (eg., their transverse and divergent waves) observed in real shipping channels and in experimental towing tanks~\cite{didenkulova13,torsvik15a,wyatt88,brown89,sheremet13,pethiyagoda18b,Ratsep20,Ratsep21,safty20,forlini21,li2021}. The theory for this line of enquiry has been almost entirely for steadily moving ships \cite{pethiyagoda17,torsvik15a,wyatt88,pethiyagoda18b}, with very brief studies of accelerating ships moving in one direction~\cite{pethiyagoda17,wyatt88}.  In this paper, we develop a new methodology to use spectrograms to analyse the signal produced by accelerating ships moving along a general path.

For steadily moving disturbances, there have been many theoretical studies that aim to predict the details of the wake, using both linear potential theory \cite{wehausen60,lighthill78} and nonlinear frameworks \cite{parau02,pethiyagoda14a,pethiyagoda14b}, with simple approximations for ship hulls (e.g.\ pressure distributions \citep{benzaquen14,darmon14,ellingsen14,miao15,moisy14a,moisy14b,pethiyagoda15,pethiyagoda21,rabaud14}) or more complicated models \cite{tuck71,noblesse13,zhu17}. In the linear regime, we can often represent the exact solution for these steady problems in integral form, with the method of stationary phase providing a nice description of the far-field, including transverse and divergent waves.
On the other hand, there are far fewer theoretical studies using linear water wave theory of unsteady ship waves.  {  For unsteady motion in a straight line, the existing studies include results for either impulsively started disturbances \cite{closa2010,wehausen60} or disturbances applied with a prescribed acceleration profile \cite{bhattarcharyya1956,doctors1972,li2019}. For vessels moving along arbitrary paths, there is a discussion in Stoker~\cite{stoker57}, for example, or notable recent work in Ref.~\cite{li2019}, where surface elevation profiles were produced for disturbances moving in a circular arc in a shear current.  The current paper adds to the small body of literature on accelerating ship waves.}

We begin our study in Sec.~\ref{sec:specSteady} by briefly reviewing the key results for computing spectrograms of steady ship wakes \cite{pethiyagoda17,pethiyagoda18b}.  In particular, by employing an example with a Gaussian pressure distribution applied to the free surface, we write down an exact solution for the surface elevation using linear potential theory and then present a spectrogram computed by taking a cross-section of the ship wake in the direction of flow (the corresponding wave signal is the same as that generated at a single point as the ship travels steadily past).  The region of high intensity in the spectrogram is shown to be very well approximated by the linear dispersion curve found using the dispersion relation.  The role of transverse and divergent waves is highlighted.

In Sec.~\ref{sec:problemDef} we consider a disturbance moving along an arbitrary path and use the method of stationary phase and geometric arguments to determine the location of the relevant dispersion curve.  While we are concerned with linear water wave theory, the derivation is written in such a way as to hold for an arbitrary dispersive medium.   In Sec.~\ref{sec:linearExamples} we provide the exact linear solution for the free-surface height at a fixed sensor due to Gaussian pressure distribution applied to the surface, moving along an arbitrary path.  This pressure distribution is a disturbance that represents a moving ship.  The ideas are illustrated via examples of spectrograms for a disturbance accelerating/decelerating in one direction and a disturbance turning in a circle.   We close that section by providing an example of an accelerating disturbance and a turning disturbance that have the same dispersion curve.  This result acts to highlight the challenges involved when attempting to uniquely identify properties of ships via surface elevation data collected at a single point. In Sec.~\ref{sec:finite} we extend the theory to apply to a finite-depth channel and then present experimental results from a towing tank to show how well the theory transfers to the laboratory.  Finally, we discuss our results in Sec.~\ref{sec:discussion}.

\section{Spectrograms of steady ship wakes}\label{sec:specSteady}

For the theoretical aspects of this paper, we are focussed on linear surface gravity waves generated by an axisymmetric pressure distribution applied to the surface and moving along some trajectory $\textbf{X}(t)=(X(t),Y(t))$.  The pressure patch  we consider is a Gaussian distribution of strength $P_0$ and characteristic length $L$.  We non-dimensionalise our problem by scaling speeds by a representative velocity $U$, lengths by $U^2/g$, and time by $U/g$, where $g$ is acceleration due to gravity.  Using these variables, the dimensionless pressure is given by $p(x-X(t),y-Y(t))$, where $p(x,y)=\epsilon\,\mathrm{exp}(-\pi^2 F^4(x^2+y^2))$ is the Gaussian, $\epsilon=P_0/\rho U^2$ is the dimensionless pressure strength, and $F=U/\sqrt{gL}$ is the Froude number. The use of applied pressure distributions to act as a simple proxy for a ship is widespread in physics \citep{benzaquen14,colen21,darmon14,ellingsen14,lo21,miao15,moisy14a,moisy14b,pethiyagoda15,pethiyagoda21,rabaud14}.  While such a basic model is unable to capture the effects of the shape of a given ship hull, it allows us to study the key features of ship waves in the time-frequency domain~\cite{pethiyagoda17,torsvik15a,wyatt88,pethiyagoda18b}, which is the goal of this work.

\begin{figure}
	\centering
	\subfloat[Free-surface]{\includegraphics[width=.5\linewidth]{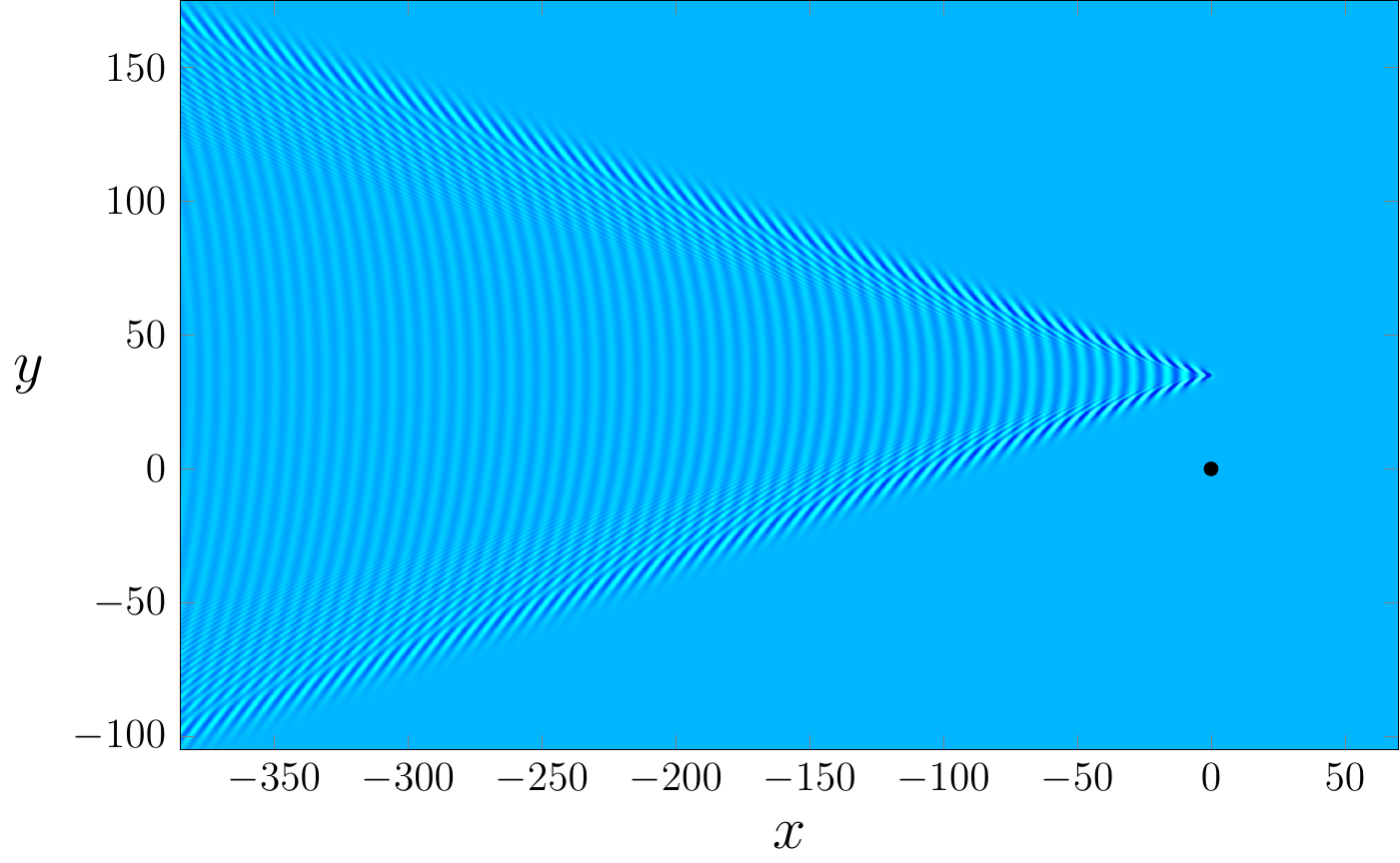}}\hspace{2ex}
	\subfloat[Spectrogram]{\includegraphics[width=.4\linewidth]{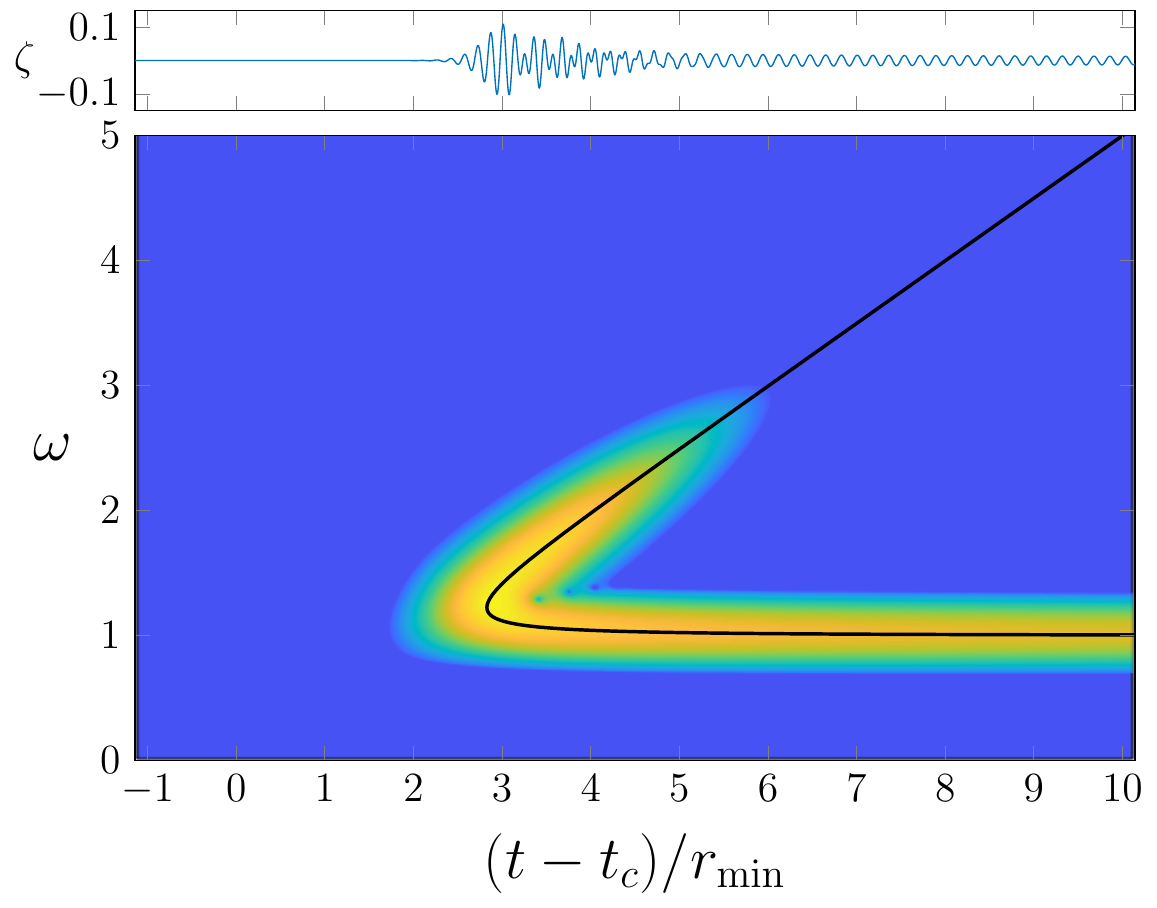}}
	\raisebox{5.6ex}{\subfloat{\includegraphics[width=.038\linewidth]{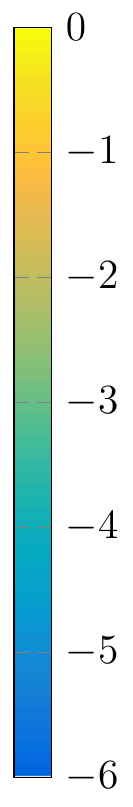}}}
	\caption{(a) A plan view of the free-surface height for steady flow past an applied Gaussian pressure distribution with $F=0.7$ at the time $t=t_c$, which is when the centre of the distribution is closest to the sensor (represented by the black dot). (b) The signal detected by the sensor placed $r_\mathrm{min}=35$ from the sailing line and its associated spectrogram.  The black curve is the classical dispersion curve which predicts the location of the highest intensity region, including the constant-frequency mode around $\omega=1$, the rising-frequency mode around $\omega=(t-t_c)/2r_{\mathrm{min}}$, and the fold where these two modes meet.}
	\label{fig:steadyExample}
\end{figure}

To illustrate the key ideas that support the application of spectrograms to analyse ship wave patterns, we begin by considering the case in which the Gaussian pressure distribution is moving steadily in one direction with unit dimensionless speed~\cite{pethiyagoda17}.  Supposing the pressure moves along the path $(X(t),Y(t))=(x_0+t,y_0)$, the dimensionless solution for the surface elevation is \citep{wehausen60}
\begin{align}
\zeta(x,y,t) = &-p(x-(x_0+t),y-y_0)+\frac{1}{2\pi^2} \int\limits_{-\pi/2}^{\pi/2}\,\int\limits_{0}^{\infty}\frac{k^2\tilde{p}(k,\psi)\cos(k[|x-(x_0+t)|\cos\psi+(y-y_0)\sin\psi])}{k-k_0}\,\,\mathrm{d}k\,\,\mathrm{d}\psi\notag\\
&+\frac{H(-x+(x_0+t))}{\pi} \int\limits_{-\pi/2}^{\pi/2}k_0^2\tilde{p}(k_0,\psi)\sin(k_0[(x-(x_0+t))\cos\psi+(y-y_0)\sin\psi])\,\,\mathrm{d}\psi,\label{eq:exactLinearInfinite}
\end{align}
where $\tilde{p}(k,\psi)=\epsilon\exp(-k^2/(4\pi^2F^4))/(\pi F^4)$ is the Fourier transform of the pressure distribution, $H(\cdot)$ is the Heaviside function and the path of integration with respect to $k$ is taken below the pole $k=k_0$, where $k_0=\sec^2\psi$.  A plan view of a wake pattern for this solution is provided in Fig.~\ref{fig:steadyExample}(a), computed for the representative value $F=0.7$.  The qualitative features of this V-shaped steady wake are well known.  For example, the transverse waves appear with crestlines that are roughly perpendicular to the direction of flow, while the divergent waves have crest lines that appear at an angle on the periphery of the wake.  A key observation is that these broad features are common to wakes behind all steadily moving disturbances, whether the disturbance is an applied pressure, a waterborne vessel, or even a duck swimming in one direction.  This is one of the reasons that the problem of deducing properties of ships from a subset of their wake is so difficult.

We now suppose there is a sensor located at the origin, a dimensionless distance $r_{\mathrm{min}}=|y_0|$ from the sailing line.  The fixed sensor measures the wave elevation as the pressure patch travels past, giving rise to the wave signal $s(t)=\zeta(0,0,t)$, which is also plotted in Fig.~\ref{fig:steadyExample}(b) (top panel) for $r_\mathrm{min}=35$. The spectrogram data for a signal $s(t)$ is given by the square magnitude of the short-time Fourier Transform
\[
S(t,\omega)=\left|\int_{-\infty}^{\infty}h(\tau-t)s(t)\mathrm{e}^{-\mathrm{i}\omega\tau}\,\mathrm{d}\tau\right|^2,
\]
where the window function, $h(t)$, is an even function with compact support. In this paper we use the Blackman-Harris 92 dB window function \cite{harris78}. All spectrograms are plotted on the scaled axis $((t-t_c)/r_\mathrm{min},\omega)$, where $t_c$ and $r_\mathrm{min}$ are the time and distance for when the ship is closest to the sensor, that is $\min(r(t))=r(t_c)=r_\mathrm{min}$, where $r(t)$ is the distance from the sensor to the ship.

The spectrogram for the signal detected by a sensor at $r_{\mathrm{min}}=35$ from the sailing line of a steadily moving applied pressure distribution is presented in Fig.~\ref{fig:steadyExample}(b) (bottom panel), where the colour scheme indicates the magnitude of the spectrogram data on the $\log_{10}$ scale. This heat map exhibits a clear region of high colour intensity that can be separated into three main parts.  The first is the constant-frequency mode (the roughly horizontal component), indicating the presence of transverse waves.  The second is the rising-frequency mode (or sliding-frequency mode, which increases with time), caused by the divergent waves.  Finally, these two modes meet at a fold, which represents the { cuspline}.

Also presented in Fig.~\ref{fig:steadyExample}(b) (bottom panel) is the linear dispersion curve (solid black curve) \cite{pethiyagoda17}.  This curve, calculated through the use of geometric arguments on wave propagation together with the dispersion relationship, predicts the frequency of a ship wave arriving at the sensor at a given time for a localised disturbance. We can see from Fig.~\ref{fig:steadyExample}(b) that the linear dispersion curve does an excellent job of indicating the location of high colour intensity in the spectrogram, and therefore provides a useful tool for associating features of spectrograms to properties of ships (or disturbances) that created the wave signal in question.  In particular, the dispersion curve successfully predicts the location of the constant-frequency mode that approaches $\omega=1$ for large $(t-t_c)/r_{\mathrm{min}}$, the rising-frequency mode that asymptotes to the line $\omega=(t-t_c)/(2r_{\mathrm{min}})$, and the fold at $((t-t_c)/r_{\mathrm{min}},\omega)=(2\sqrt{2},\sqrt{6}/2)$, corresponding to the well known Kelvin wake angle $\mathrm{arctan}\,(1/\sqrt{8})$.  One of the key outcomes of the present study is to extend this linear dispersion curve to examples for which the ship is no longer moving with constant velocity.

\section{Dispersion curves for unsteady motion}\label{sec:problemDef}

In this section we develop our theoretical results for disturbances moving along arbitrary paths.  In Sec.~\ref{sec:problemDef_I} we present the exact solution for the surface elevation due to moving disturbance.  This solution allows us to simulate wave elevation signals measured at a single point and visualise the wave frequencies via a spectrogram.  In Sec.~\ref{sec:problemDef_II} we apply the method of stationary phase to the exact solution to derive the classical dispersion curve, which predicts the location of the highest intensity colour in the spectrogram.  We also consider end-point contributions to the integral, which lead to curves we call start-point and final-point dispersion curves.  Alternative arguments to derive these dispersion curves are provided in Sec.~\ref{sec:problemDef_III}; these arguments are based in part on preliminary ideas outlined in Ref.~\cite{pethiyagoda17}.

\subsection{Signal produced by a pressure moving along an arbitrary path}\label{sec:problemDef_I}

In dimensionless variables, suppose we have a stationary sensor at the origin and a moving disturbance whose relative position to the sensor and velocity are given by $\textbf{X}(t)$ and $\textbf{X}^\prime(t)=\textbf{U}(t)$, respectively. The associated ship speed is given by $U(t)=|\textbf{U}(t)|$, which is of order one.  As with any dispersive medium, we can define the dispersion relation, phase and group velocities as
\begin{align}
\omega=\Omega(k),\qquad
c_p=\frac{\Omega(k)}{k},\qquad
c_g=\frac{\mathrm{d}\Omega(k)}{\mathrm{d}k} \label{eq:disp},
\end{align}
respectively, where $\omega$ is the wave frequency, $\Omega(k)$ is the wave function, and $k$ is the wavenumber.  Our focus in this paper is on linear water wave theory, for which $\Omega(k)$ will depend on whether we assume infinitely deep water ($\Omega(k)=\sqrt{k}$) or take into account a bottom topography (for a channel of constant depth, $\Omega(k)=\sqrt{k\tanh(k/F_H^2)}$, where $F_H=U/\sqrt{gH}$ is the depth-based Froude number and $H$ is the dimensional water depth).  However, we shall keep $\Omega(k)$ general in this section to cover these and other possibilities.

{ Following Ref.~\cite{stoker57},} we construct the surface elevation of a moving disturbance to be the superposition of pressure pulses applied along the ship's path $\textbf{X}(t)$, giving
\begin{align}
\zeta(\textbf{x},t)=&\int_0^{t}\zeta_p(r(\textbf{x},\tau),\theta(\textbf{x},\tau),t-\tau)\,\mathrm{d}\tau,\label{eq:ogsignal}
\end{align}
where
\begin{equation}
\zeta_p(r,\theta,t)=-\frac{1}{4\pi^2}\int_0^\infty\int_{-\pi}^\pi \frac{k^2}{\Omega(k)}\tilde{p}(k,\psi)\sin\left(\Omega(k)t\right)\mathrm{e}^{\mathrm{i}kr\cos(\theta-\psi)}\,\mathrm{d}\psi\,\mathrm{d}k,\label{eq:zetap}
\end{equation}
is the surface elevation for a pressure impulse in polar coordinates ($x=r\cos\theta, y=r\sin\theta$), $r(\textbf{x},\tau)=|\textbf{X}(\tau)-\textbf{x}|$ is the radial distance between the centre of the pressure and a point on the surface, $\theta(\textbf{x},\tau)= \cos^{-1}({-(\textbf{X}-\textbf{x})\cdot\textbf{U}}/{rU})$ is chosen under the assumption that the ``front'' of the pressure distribution, $p(x,y)$, is located on the positive $x$-axis and will be orientated to point in the direction of travel when moving along the path. { For completeness the derivation of (\ref{eq:zetap}) is given in Appendix \ref{sec:derivingGovEq}.} The above integral can be used to generate a wave signal cause by an impulsively switched-on pressure distribution by evaluating (\ref{eq:ogsignal}) at the origin $s(t)=\zeta(\textbf{0},t)$, where the sensor is situated.

\subsection{Method of stationary phase}\label{sec:problemDef_II}

To construct the linear dispersion curve as a multivalued function of the horizontal time axis of the spectrogram, $t$, we determine the important frequency values $\omega$ for a given value of $t$. We begin by rearranging the signal function (\ref{eq:ogsignal}) by separating the oscillating and non-oscillating components,
\begin{equation}
s(t)=\int_0^{t}\int_0^\infty\int_{-\pi}^\pi A(k,\psi)\left(\mathrm{e}^{\mathrm{i}g_1(k,\psi,\tau)}-\mathrm{e}^{\mathrm{i}g_2(k,\psi,\tau)}\right)
\,\mathrm{d}\psi\,\mathrm{d}k\,\mathrm{d}\tau,\label{eq:sepsignal}
\end{equation}
where $A(k,\psi)$ is the non-oscillating amplitude function and $g_{1,2}(k,\psi,\tau)=kr(\tau)\cos(\theta(\tau)-\psi)\pm\Omega(k)(t-\tau)$ is the phase function. Note we are simplifying our notation by setting $r(\tau)=r({\bf 0},\tau)$, $\theta(\tau)=\theta({\bf 0},\tau)$, since we are concerned with measuring our signal at the sensor, which we have fixed to be at the origin.

{
For a given path of the ship, there will be a characteristic (dimensional) acceleration, $a$ say, which will lead to a dimensionless parameter $\alpha=a/g$.  For the stipulated velocity of the ship to be dimensionally correct, it must be of the form $\textbf{U}(t)= \mathrm{function}\,(\alpha t)$, where the function is an order one quantity.  Assuming that $\alpha\ll 1$, which is the physically realistic limit, then this ansatz for the velocity implies the position vector (and therefore $r(t)$) is $\mathcal{O}(1/\alpha)$.  Therefore, by introducing a shorter time-scale $\tilde{t}=\alpha t$ in (\ref{eq:sepsignal}), we see the integral is in the appropriate form for the method of stationary phase, provided $\alpha\ll 1$.  In practice the method appears to work quite well even when $\alpha$ is not small.
}

Following the methodology of stationary phase for multiple integrals \cite{stoker57,jones58}, the main contribution to the integral (\ref{eq:sepsignal}) occurs at the stationary points, namely when all three partial derivatives of $g_{1,2}$ vanish.  The relevant partial derivatives are
\begin{align}
\frac{\partial g_{1,2}}{\partial \psi}&=kr(\tau)\sin(\theta(\tau)-\psi)  = 0, \label{eq:psidiff}\\
\frac{\partial g_{1,2}}{\partial k}&=r(\tau)\cos(\theta(\tau)-\psi)\pm\Omega^\prime(k)(t-\tau) = 0,\label{eq:kdiff}\\
\frac{\partial g_{1,2}}{\partial \tau}&=kr^\prime(\tau)\cos(\theta(\tau)-\psi)- kr(\tau)\sin(\theta(\tau)-\psi)\theta^\prime(\tau)\mp\Omega(k) = 0,\label{eq:taudiff}
\end{align}
where the prime indicates a derivative of the function with regard to the argument. { Note we are only interested in the stationary points and do not require the approximation itself.} From Eq.~(\ref{eq:psidiff}) we have $\psi=\theta(\tau)$ which, when substituted into (\ref{eq:kdiff})-(\ref{eq:taudiff}), gives the relations
\begin{equation}
r(\tau)\pm\Omega^\prime(k)(t-\tau) = 0,\qquad kr^\prime(\tau)\mp\Omega(k) = 0,\label{eq:diffrelmid}
\end{equation}
respectively. Equations (\ref{eq:diffrelmid}) can be rearranged and further simplified by assuming that we are only interested in positive group and phase velocities,
\begin{align}
c_g(k)&=\frac{r(\tau)}{t-\tau},\label{eq:kImpicit}\\
c_p(k)&=-r^\prime(\tau),\label{eq:cpPhys}
\end{align}
where $c_g$ is the group velocity and $c_p$ is the phase velocity. Equations (\ref{eq:kImpicit})-(\ref{eq:cpPhys}) can be taken simultaneously to give solutions of interest $\tau=\tau_j$ and $k=k_j$ for $j=1,\ldots,n$ (for surface gravity waves, $n=0$, $1$ or $2$, depending on whether the sensor is situation outside, on, or inside the caustics at a given time).  The values $k_j$ are then used to give the frequency location, $\omega$, of colour intensity on the spectrogram using the dispersion function (\ref{eq:disp}). The curves generated by these solutions of interest are the unsteady analogue to the steady dispersion curve derived in Ref.~\cite{pethiyagoda17} (and presented in Fig.~\ref{fig:steadyExample}(b)).
	
We consider two additional contributions of interest, namely those from the start time $t=0$ and the final time $t=t_f$.  According to the method of stationary phase, these end-point contributions are formally smaller than those associated with the stationary points (where partial derivatives of $g_{1,2}$ vanish), however in certain instances (for example, when the dimensionless measure of acceleration is large) they help explain unexpected features in the time-frequency domain.  For these end-point contributions, we say they occur at $t=\tau_e$, so that at the start and final times, we have $\tau=\tau_e=0$ and $\tau=\tau_e=t_f$, respectively; in both cases, the relevant $k$ is given implicitly through (\ref{eq:kImpicit}) and the frequency is then computed via by the dispersion function (\ref{eq:disp}).  These contributions therefore correspond to frequencies $\omega$ as a function of time that can be plotted on top of the spectrograms.  The two such curves on each spectrogram will be referred to as the start-point and final-point dispersion curves, respectively (these will be coloured red in our figures).  Note that if the ship is still moving at $t=t_f$ then the final-point dispersion curve's frequency is $\omega=0$.

It is worth clarifying the physical interpretation of these start-point and final-point contributions.  First, the start-point contribution is equivalent to that due to a single pulse of a pressure distribution $p$, centred at $\textbf{X}(0)$, applied at $t=0$, which emits waves of every wave number, each moving at a wave-number-dependent speed.  Low frequency waves that travel faster than the ship will arrive at the sensor before the actual wake does.  For ships moving with large acceleration, these frequencies are picked up by spectrograms.  Similarly, the final-point contribution is equivalent to a negative pulse of a pressure distribution centred at $\textbf{X}(t_f)$, applied at $t=t_f$; for gravity waves propagating on an infinitely deep fluid, this contribution gives rise to a dispersion curve that is a line with slope $1/2$.  For this final-point contribution, the frequencies may be observable in the time-frequency domain for decelerating ships, especially if the rate of deceleration is high.


\subsection{Geometric arguments}\label{sec:problemDef_III}

\begin{figure}
	\centering
	\includegraphics[width=.6\linewidth]{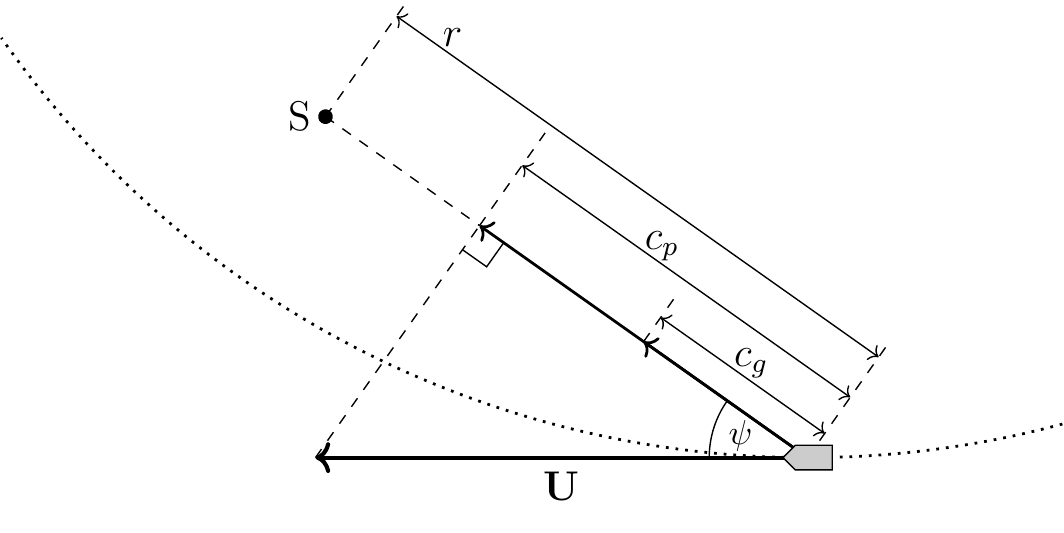}
	\caption{A schematic of a ship (grey object) travelling along an arbitrary path (dotted curve) with velocity $\textbf{U}$. A sensor is located at S where $r$ is the radial distance from sensor to the ship and $\psi$ is the angle between the sailing line and the direction to the sensor. Waves generated by the ship travel with phase velocity $c_p$ and group velocity $c_g$. The right angle triangle between $\textbf{U}$ and $c_p$ is a consequence of (\ref{eq:cpPhys}).}
	\label{fig:scheme}
\end{figure}

It is possible to use geometry to interpret the different dispersion curves in terms of physical properties for a better understanding of how the disturbance influences the wave elevation signal.  By referring to the schematic in Fig.~\ref{fig:scheme} we infer that a ship moving along an arbitrary path (dotted curve) with velocity $\textbf{U}$ will generate wave packets travelling with group velocity $c_g$\cite{stoker57}. We can define a dispersion curve parametrically by assuming that the time a wave packet arrives at the sensor, $t$, is given by the time the wave packet was generated, $\tau$, plus the time taken for the wave packet to travel from the ship to the sensor.  The frequency of the wave packet is then given by the dispersion relation (\ref{eq:disp}). That is, each dispersion curve is given by
\begin{equation}
(t,\omega)=\left(\tau+\frac{r(\tau)}{c_g(k)},\Omega(k)\right),
\end{equation}
where the time coordinate is equivalent to (\ref{eq:kImpicit}).

In order to derive the classical dispersion curve, we rewrite (\ref{eq:cpPhys}) as
\begin{equation}
c_p=U\cos(\psi),\label{eq:cpTheta}
\end{equation}
where
\begin{equation}
\psi = \cos^{-1}\left(\frac{-\textbf{X}\cdot\textbf{U}}{rU}\right)\label{eq:theta}
\end{equation}
is the angle between the sailing line and direction to the sensor (Fig.~\ref{fig:scheme}). We can then use (\ref{eq:cpTheta}) to define the in-phase wavenumber $k_j$ as a function of the wave packet generation time $\tau$ and, therefore, the classical dispersion curve can be defined as
\begin{equation}
(t(\tau),\omega(\tau))=\left(\tau+\frac{r(\tau)}{c_g(k_j(\tau))},\Omega(k_j(\tau))\right)\quad \text{for } 0\leq\tau\leq t.\label{eq:paraDisp}
\end{equation}

Now turning to the additional dispersion curves, we recall that the start-point and final-point contributions can be thought to be due to a pulse disturbance (that is, a disturbance that produces waves of all wavenumbers $k$ at a single instance) at the ship's location when it starts and finishes its movement.  As such, the start-point and final-point dispersion curves can be expressed parametrically by
\begin{equation}
(t(k),\omega(k))=\left(\tau_e+\frac{r(\tau_e)}{c_g(k)},\Omega(k)\right),\quad \text{for }k\ge 0,\label{eq:pulseDisp}
\end{equation}
where here $\tau_e=0$ and $\tau_e=t_f$, respectively.

\section{Accelerating ships on an infinitely deep body of water}\label{sec:linearExamples}

For the linear ship wave examples we consider here, we apply the axisymmetric pressure distribution, $p(x,y)=\epsilon\,\exp(-\pi^2F^4(x^2+y^2))$, whose Fourier transform is given by $\tilde{p}(k)=\epsilon\exp(-k^2/4\pi^2F^4)/(\pi F^4)$. Simplifying (\ref{eq:zetap}) for an axisymmetric pressure distribution gives
\begin{equation}
\zeta_p(r,t)=-\frac{1}{2\pi}\int_0^\infty \frac{k^2\tilde{p}(k)}{\Omega(k)}\sin\left(\Omega(k)t\right)J_0(kr)\,\mathrm{d}k,
\end{equation}
where $J_0(x)$ is the Bessel function of the first kind of order zero.

In order to remove the effect of impulsively switching on the pressure distribution we assume the pressure distribution has been stationary at its starting location for all time $t<0$. The equations for this initial disturbance are
\begin{align}
\frac{\mathrm{d}^2\tilde{\zeta}_i}{\mathrm{d}t^2}+\Omega(k)^2\tilde{\zeta}_i = 0,\label{eq:zetaDODE}\\
\tilde{\zeta}_i|_{t=0}=-\frac{k\tilde{p}(k)}{\Omega(k)^2},\label{eq:zetaInitD}\\
\frac{\mathrm{d}\tilde{\zeta}_i}{\mathrm{d}t}\Big|_{t=0} = 0,\label{eq:zetaDinit}
\end{align}
where $\tilde{\zeta}_i$ is the Fourier transform of the surface elevation due to the initial disturbance. Solving (\ref{eq:zetaDODE})-(\ref{eq:zetaDinit}) and inverting the Fourier transform gives
\begin{equation}
\zeta_i(r,t)=-\frac{1}{2\pi}\int_0^\infty \frac{k^2\tilde{p}(k)}{\Omega(k)^2}\cos\left(\Omega(k)t\right)J_0(kr)\,\mathrm{d}k.
\end{equation}
We can now consider the surface elevation of a moving ship to be the superposition of the initial disturbance and the pressure pulses applied along the ships path $\textbf{X}(t)$, giving
\begin{equation}
\zeta(\textbf{x},t)=\zeta_i(r(\textbf{x},0),t)+\int_0^{t}\zeta_p(r(\textbf{x},\tau),t-\tau)\,\mathrm{d}\tau,\label{eq:surface}
\end{equation}
recalling that the signal at the sensor is given by $s(t)=\zeta(\textbf{0},t)$ where $r(\textbf{x},t)=|\textbf{X}(t)-\textbf{x}|$. For the remainder Sec.~\ref{sec:linearExamples} we will only study infinite-depth gravity waves so that $\Omega(k)=\sqrt{k}$ (finite-depth examples will be considered in Sec.~\ref{sec:finite}).

\subsection{Ship accelerating in a straight line}\label{sec:accStraight}

\begin{figure}
	\begin{tabular}{cc}
		\multicolumn{2}{c}{\raisebox{20ex}{\footnotesize(a) $t=t_c$\hspace{3.5em}}\subfloat{\includegraphics[width=.65\linewidth]{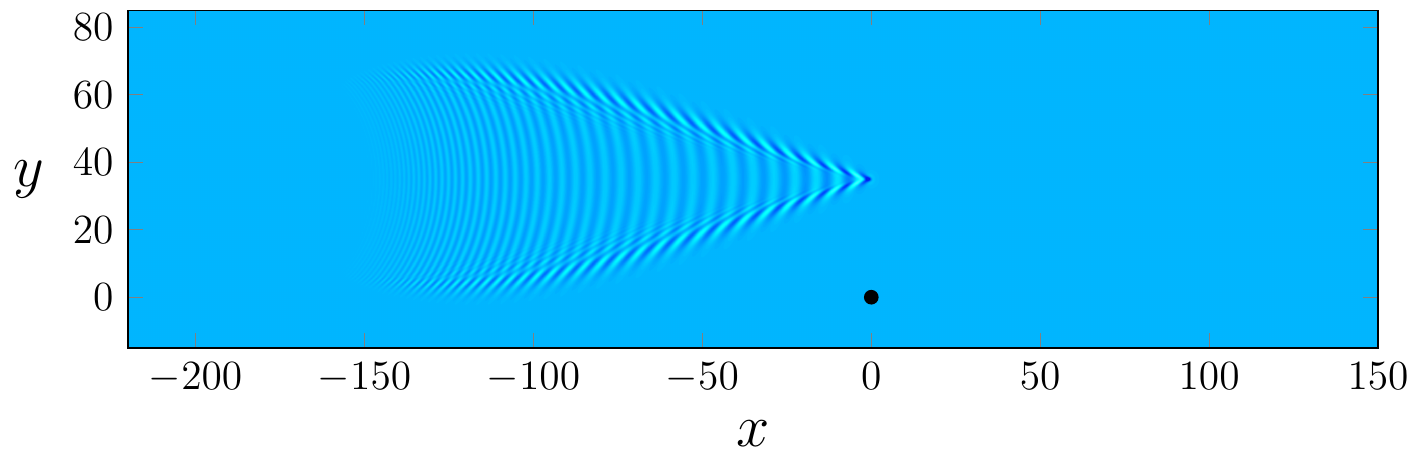}}}\\
		\multicolumn{2}{c}{\raisebox{20ex}{\footnotesize(b) $t=t_c+4r_\mathrm{min}$}\subfloat{\includegraphics[width=.65\linewidth]{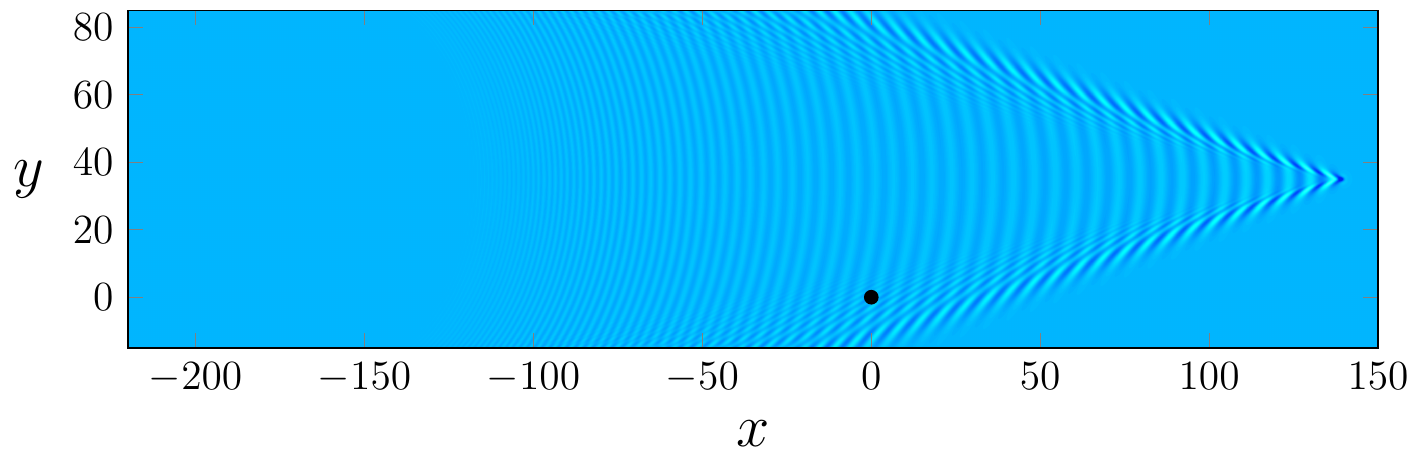}}}\\
		\subfloat[$\alpha=\infty$]{\includegraphics[width=.42\linewidth]{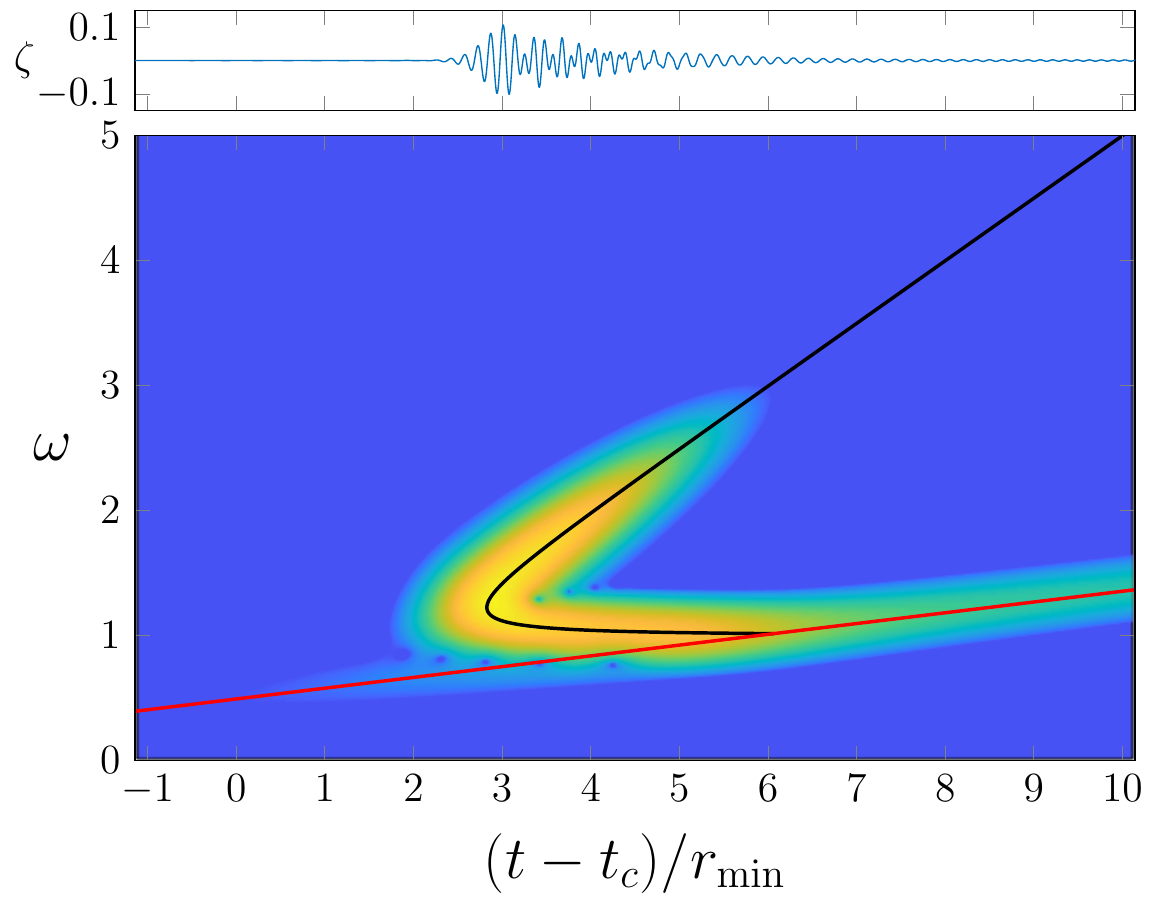}}\hspace{2ex}
		\subfloat[$\alpha = 0.25$]{\includegraphics[width=.42\linewidth]{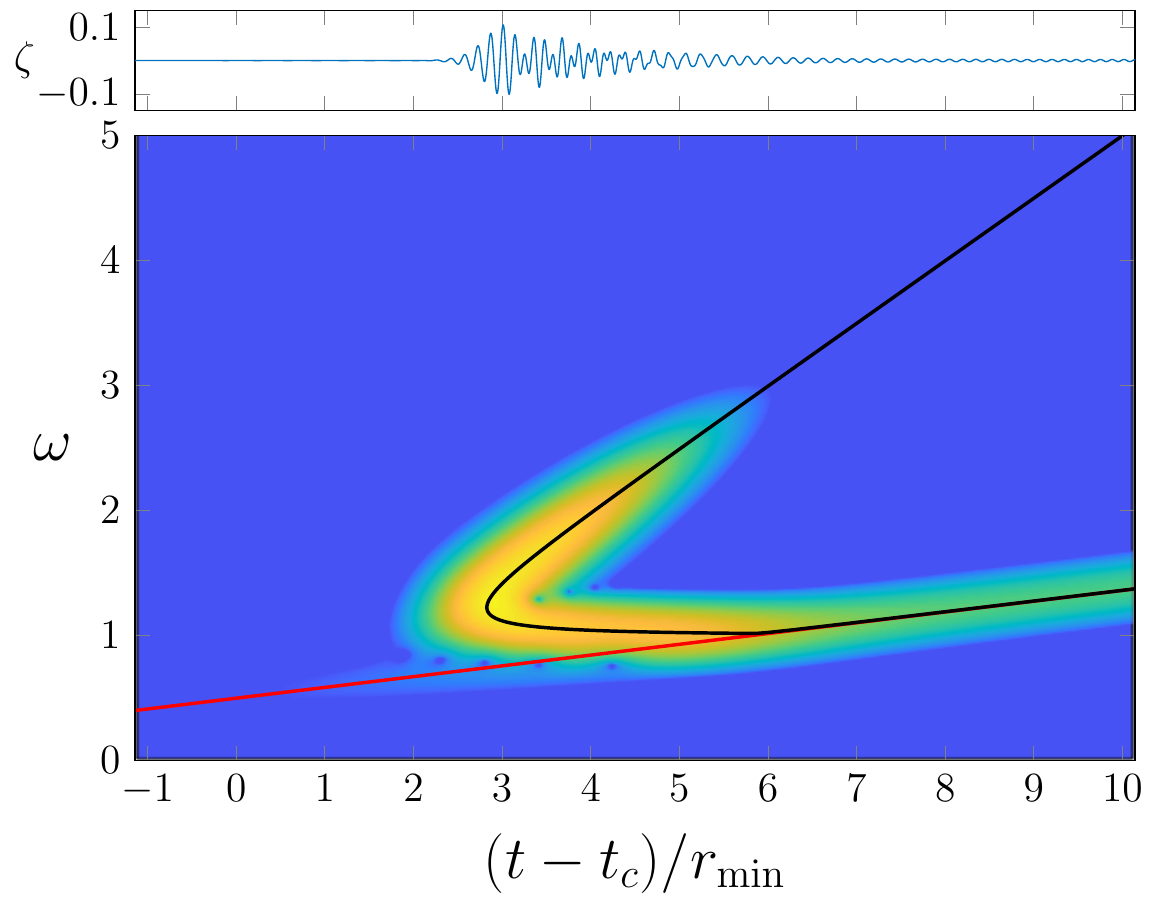}}&
		\hspace{-.0\linewidth}\multirow{4}{*}[0.17\linewidth]{\includegraphics[width=.06\linewidth]{colourbarLinear.pdf}}\\
		\subfloat[$\alpha=0.1$]{\includegraphics[width=.42\linewidth]{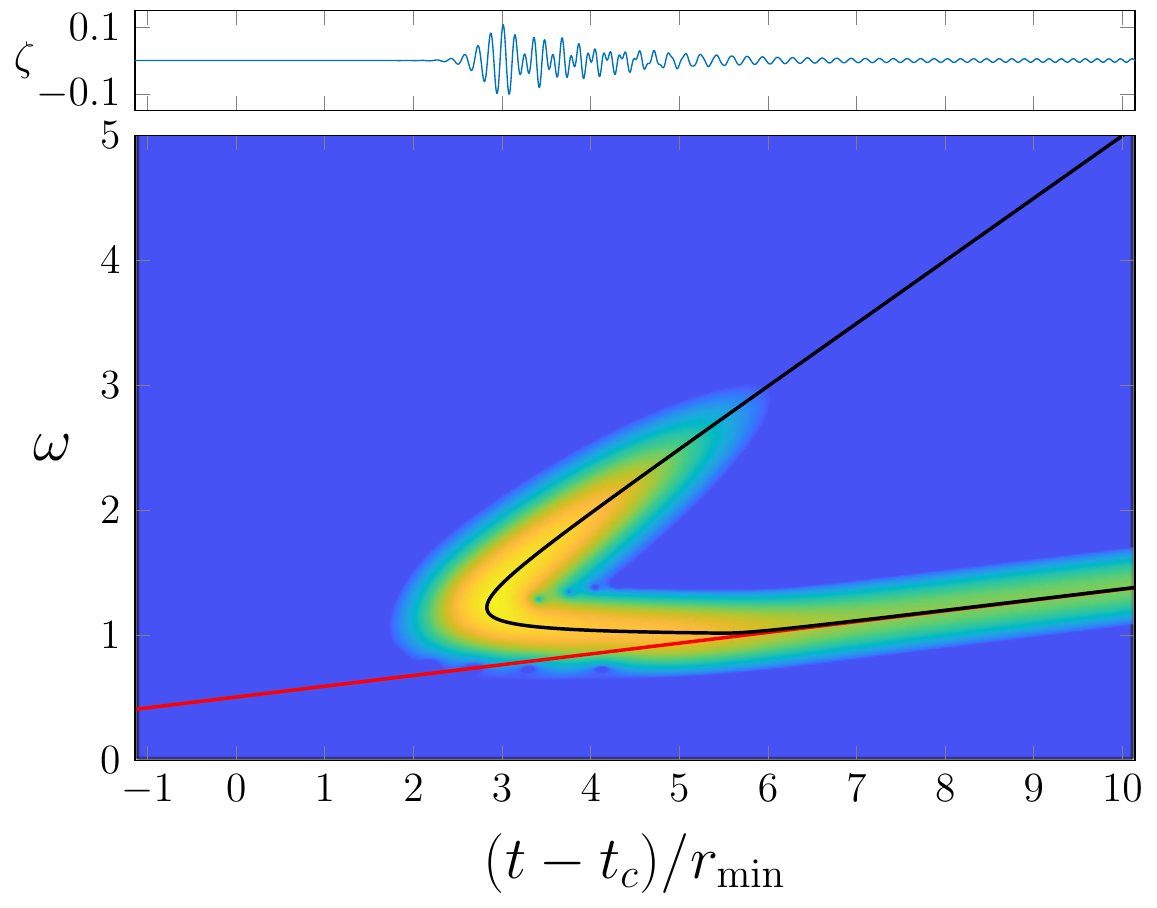}}\hspace{2ex}
		\subfloat[$\alpha=0.01$]{\includegraphics[width=.42\linewidth]{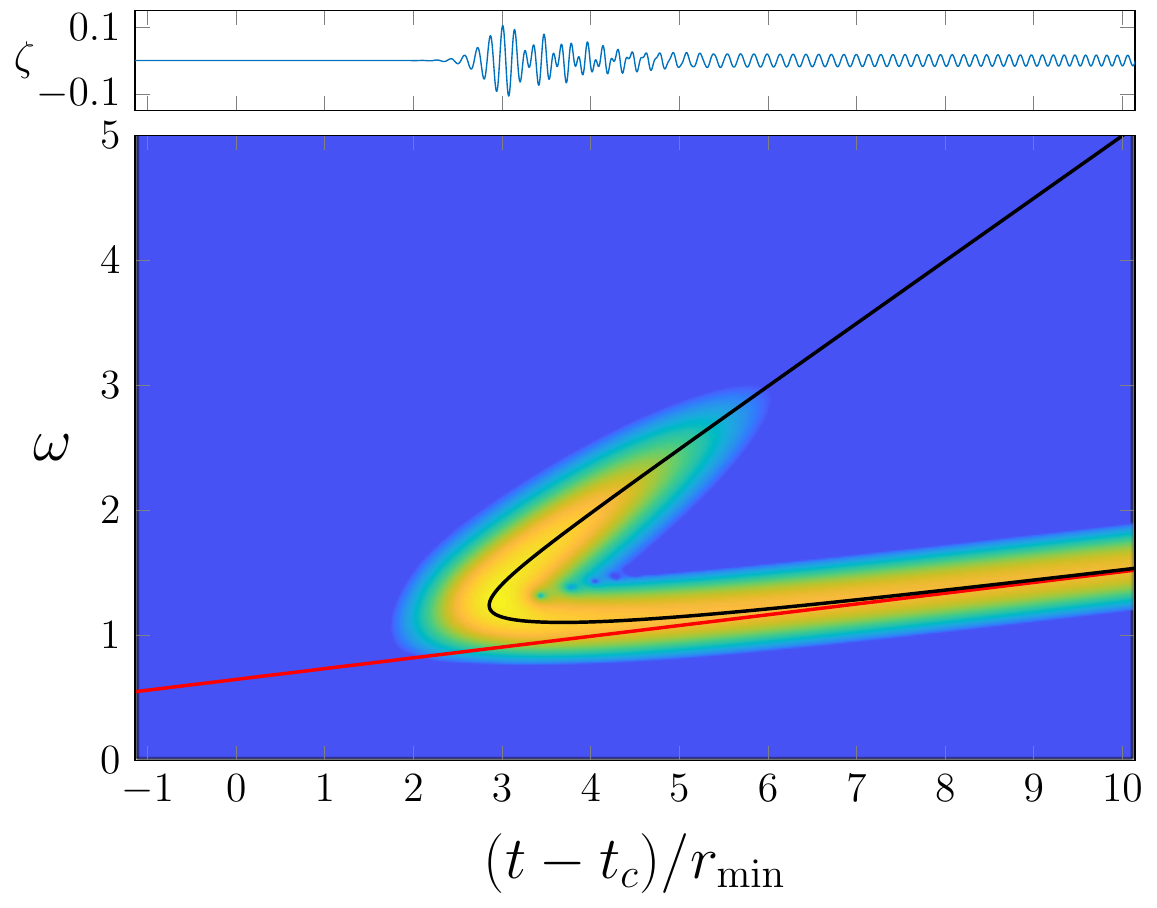}}&\\
	\end{tabular}
	\caption{(a,b) Shows the free-surface height, evaluated using (\ref{eq:surface}), for a pressure distribution with $F=0.7$ starting at the point $(-200,35)$ and moving with speed profile (\ref{eq:shipSpeed}) for $\alpha=0.01$ at two different times $t=t_c$ and $t_c+4r_\mathrm{min}$, the black dot is the location of the sensor. (c--f) Spectrograms computed for acceleration $\alpha=\infty,0.25,0.1,0.01$ (note: $\alpha=\infty$ refers to an impulsively moved ship). The classical (black) and start-point (red) dispersion curves are overlaid. }
	\label{fig:LinSpec}
\end{figure}

The first example we will consider is a ship accelerating in a straight line represented by
\begin{equation}
\textbf{X}(t)=\left(x_0+\int_0^t u(\tau)\,\mathrm{d}\tau,y_0\right),\qquad \textbf{U}(t)=(u(t),0),\label{eq:straightPath}
\end{equation}
where the ship's starting location is positioned at the point $(x_0,y_0)$ relative to the sensor, and $u(t)$ is the ship's speed in the positive $x$-direction given by
\begin{equation}
u(t)=\mathrm{erf}\left(\frac{\sqrt{\pi}\alpha t}{2}\right),\label{eq:shipSpeed}
\end{equation}
where $\alpha$ is the initial dimensionless acceleration.  As an example, wave patterns for disturbances travelling with velocity (\ref{eq:shipSpeed}) are shown in Fig.~\ref{fig:LinSpec}(a,b).  Here $F=0.7$ and $\alpha=0.01$.  These plots show that, even though the disturbance is accelerating, the wake still exhibits the characteristic transverse and divergent waves that are observed for steady ships, with shorter wavelengths towards the end of the wake.  The effect of acceleration is to bend the edges of the V-shaped pattern towards the centreline.

Figure \ref{fig:LinSpec} shows the spectrogram produced by the ship for $\alpha=0.01,0.1,0.25,\infty$ ($\alpha=\infty$ refers to an impulsively accelerated ship), where the (black) classical and (red) start-point dispersion curves are overlaid. For this example, the final-point dispersion curve has a frequency of zero and so is not present.  The classical dispersion curve is comprised of two branches in the same way as the uniform-velocity version presented in Fig.~\ref{fig:steadyExample}(b).  Following the terminology used with steady ship wakes \cite{pethiyagoda17}, we refer to the lower (upper) branch of the classical dispersion curve as the transverse (divergent) branch due to its relationship with the transverse (divergent) waves. Interestingly, if we were to attempt an analogous classification here and define a transverse wave as being associated with $\psi<\tan^{-1}(1/\sqrt{2})$ and divergent waves with $\psi>\tan^{-1}(1/\sqrt{2})$, where $\psi$ is given by (\ref{eq:theta}), then for the accelerating disturbance presented in this section the borderline between transverse and divergent waves occurs below the fold towards the beginning of the lower branch (recall that for steady waves this delineation point occurs precisely at the fold itself).  In fact, for an accelerating disturbance, the borderline propagation direction $\psi$ and wavenumber $k$ depend heavily on the path taken, so the distinction between transverse and divergent waves is not as clear.

From Fig.~\ref{fig:LinSpec} we can see that, in all cases, the divergent branch approaches the line $\omega=(t-t_c)/2r_\mathrm{min}$ in the same way as it does for a steadily moving disturbance \cite{pethiyagoda17}; given a spectrogram computed from experimental data, this branch can be used to estimate the values of $t_c$ and $y_0$.  For an impulsively moved ship (Fig.~\ref{fig:LinSpec}(c)) the transverse branch roughly follows the constant frequency mode $\omega=1$, but is truncated when it touches the start-point dispersion curve; this is because any waves along the transverse branch for later times would have been produced by the ship before it started moving. Alternatively, when a ship has finite acceleration the transverse branch increases in frequency as time increases, asymptotically approaching the start-point dispersion curve, given for infinite depth by $\omega=t/2r(0)$. As with the divergent branch, the start-point dispersion curve can be used to estimate the initial location of the ship $(x_0,y_0)$ (recall $y_0$ is given by the divergent branch).

When using the spectrogram to estimate the speed of a steadily moving ship, Pethiyagoda \emph{et al.} \cite{pethiyagoda17} made use of the property that the transverse branch asymptotically approaches $\omega=1$ for disturbances moving at constant velocity. Unfortunately, in the case of an accelerating ship, the asymptote is dictated by the ship's initial position and cannot be directly used to estimate the ship's speed. However, the minimum frequency of the classical dispersion curve can be used as a lower bound on the ship's speed, where the farther along the transverse branch the minimum occurs the closer the lower bound is to the true ship speed. To approximate the full velocity profile, the parametric form of the dispersion curve must be sampled to form a system of equations given by (\ref{eq:paraDisp}) where the speed $u(\tau)$ is calculated using a finite difference approximation and the unknowns are time $\tau$ and the $x$-position of the ship.

Focusing on the colour intensity of the spectrograms in Fig.~\ref{fig:LinSpec}, we can see that the classical dispersion curve accurately predicts the location of colour intensity for all examples. The start-point dispersion curve indicates another region of colour intensity that is more prominent for higher values of the acceleration $\alpha$.  As such, this example demonstrates the effectiveness of the theory behind the dispersion curves, especially when acceleration is important.

\subsection{Ship decelerating in a straight line}

\begin{figure}
	\begin{tabular}{cc}
		\multicolumn{2}{c}{\raisebox{20ex}{\footnotesize(a) $t=t_c$\hspace{3.5em}}\subfloat{\includegraphics[width=.65\linewidth]{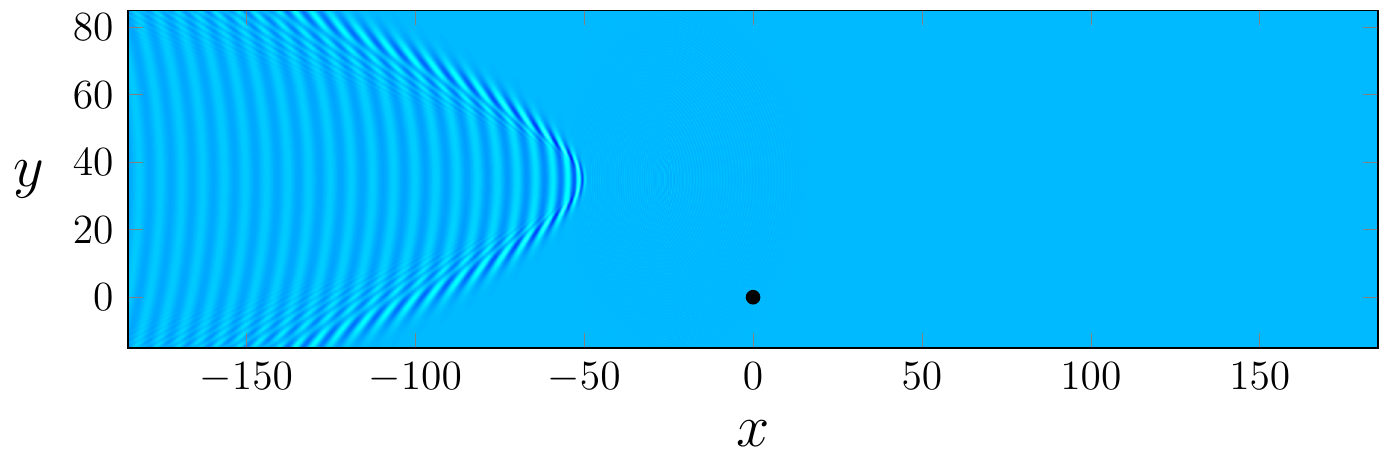}}}\\
		\multicolumn{2}{c}{\raisebox{20ex}{\footnotesize(b) $t=t_c+4r_\mathrm{min}$}\subfloat{\includegraphics[width=.65\linewidth]{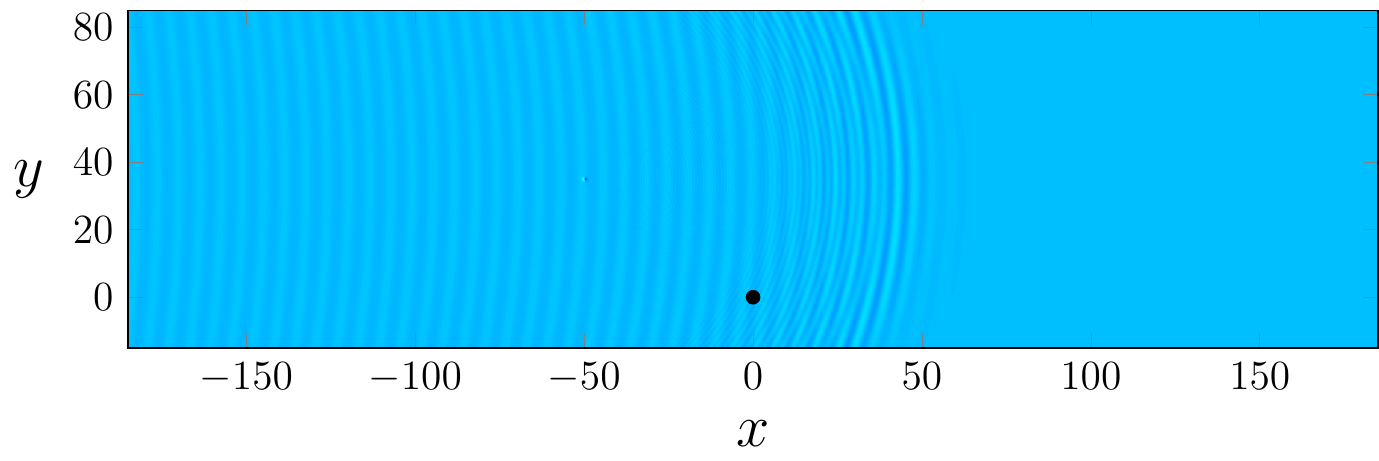}}}\\
		\subfloat[$\alpha=-\infty$]{\includegraphics[width=.42\linewidth]{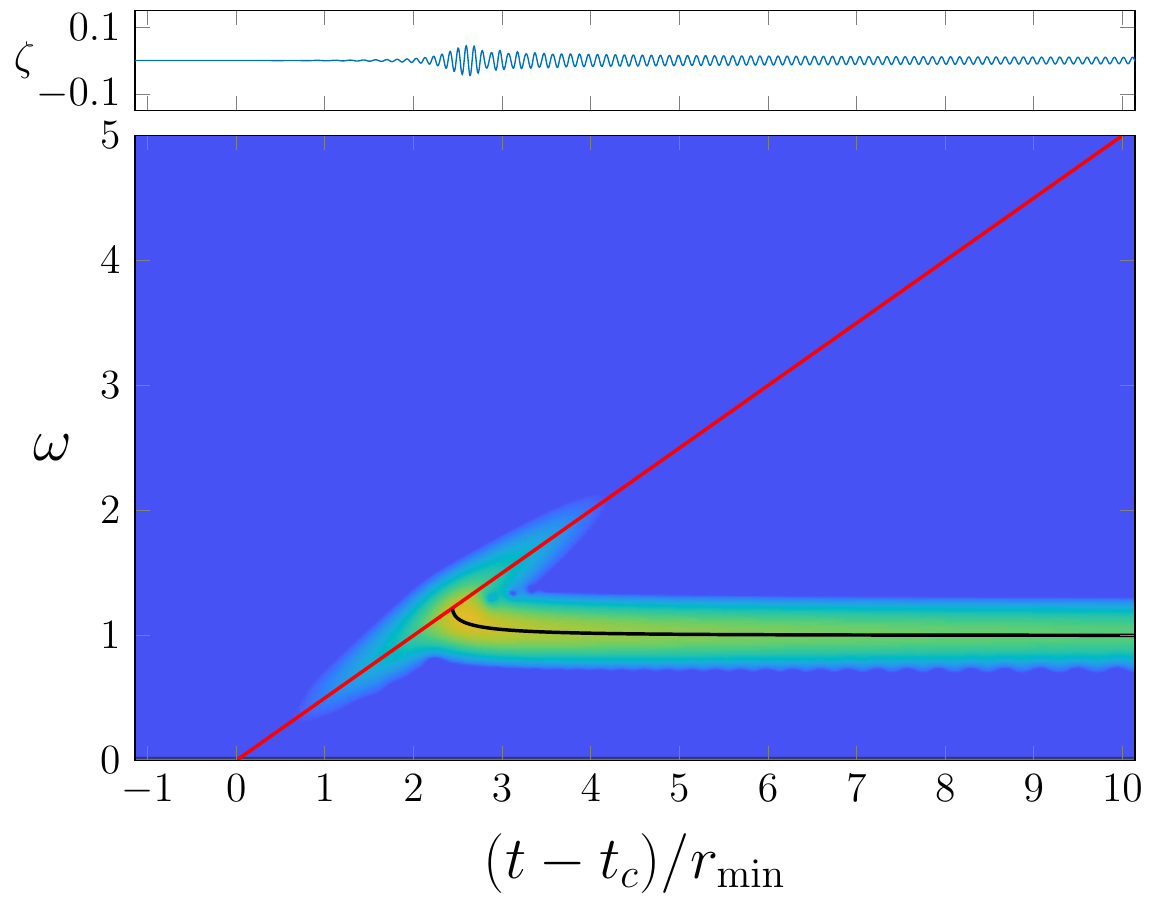}}\hspace{2ex}
		\subfloat[$\alpha = -0.25$]{\includegraphics[width=.42\linewidth]{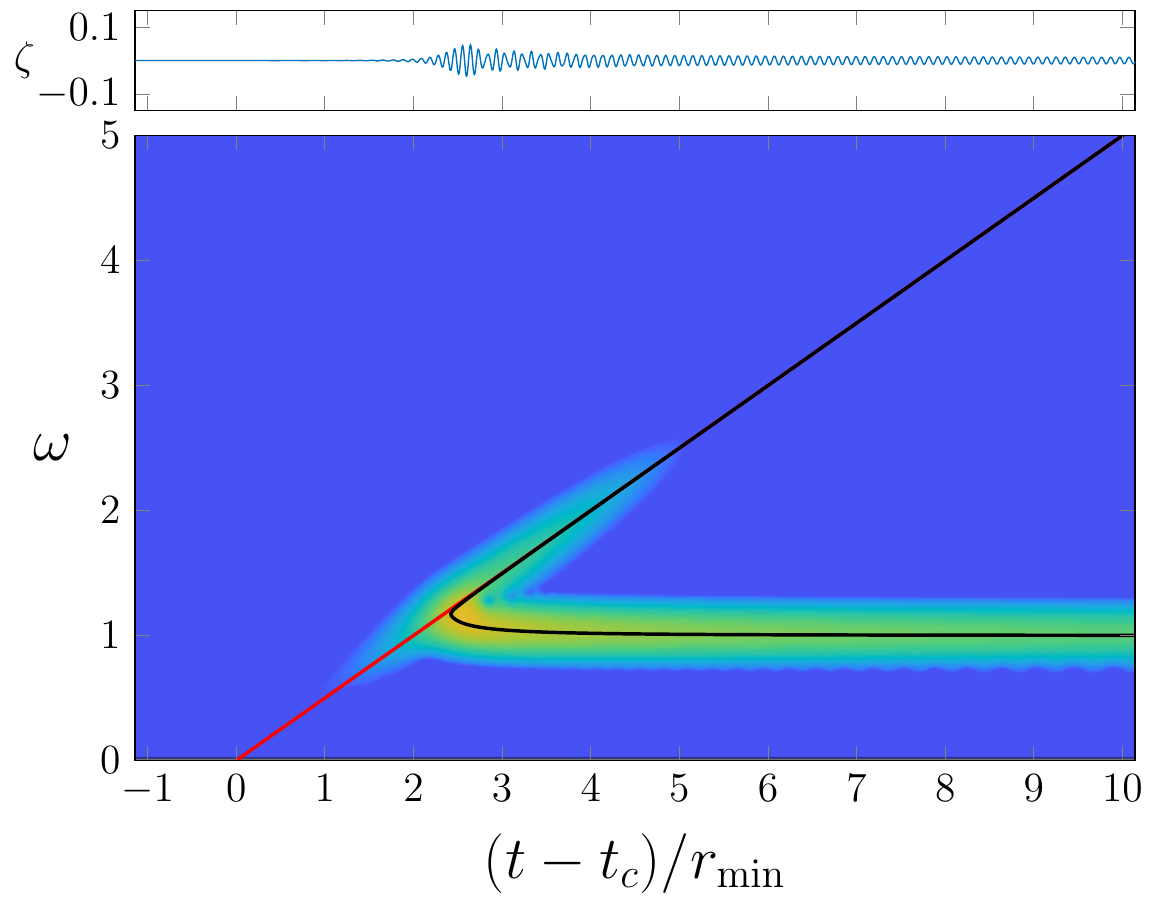}}&
		\hspace{-.0\linewidth}\multirow{4}{*}[0.17\linewidth]{\includegraphics[width=.06\linewidth]{colourbarLinear.pdf}}\\
		\subfloat[$\alpha=-0.1$]{\includegraphics[width=.42\linewidth]{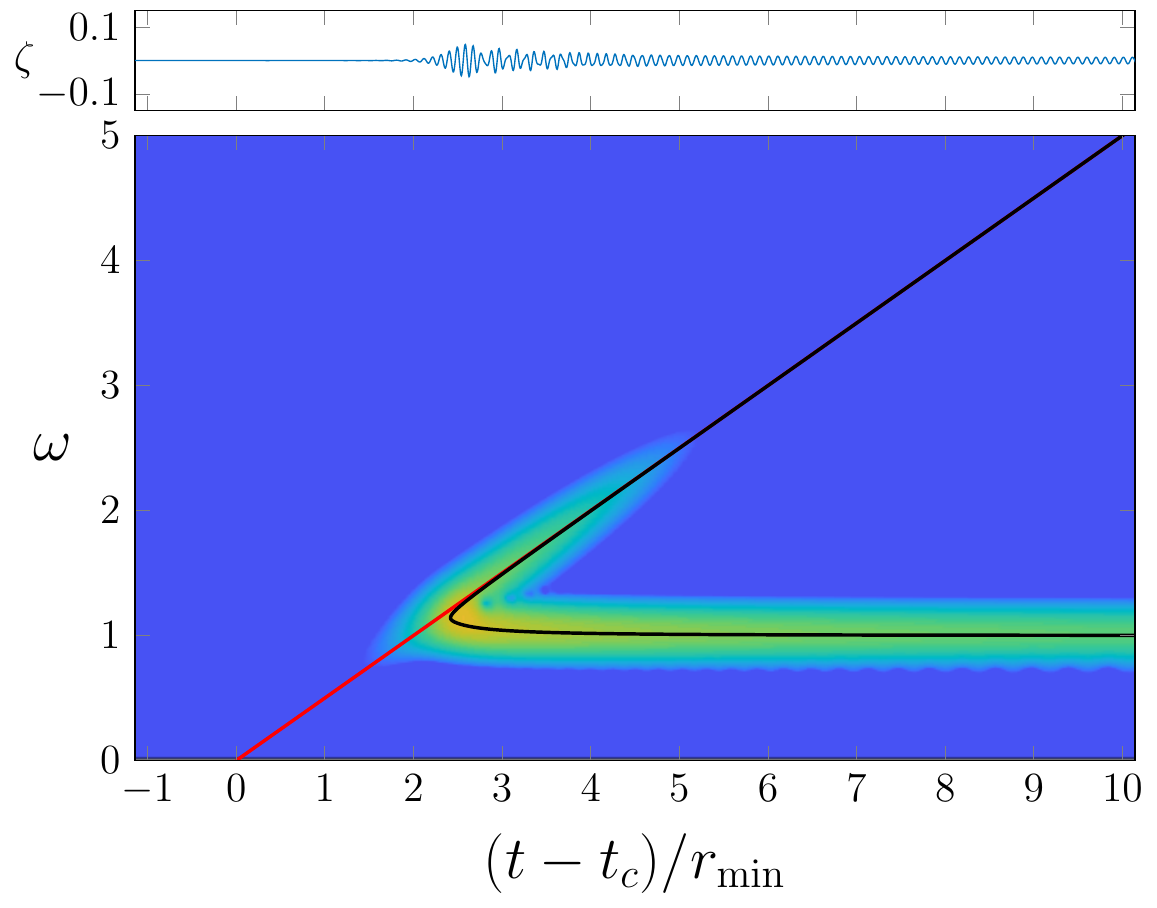}}\hspace{2ex}
		\subfloat[$\alpha=-0.01$]{\includegraphics[width=.42\linewidth]{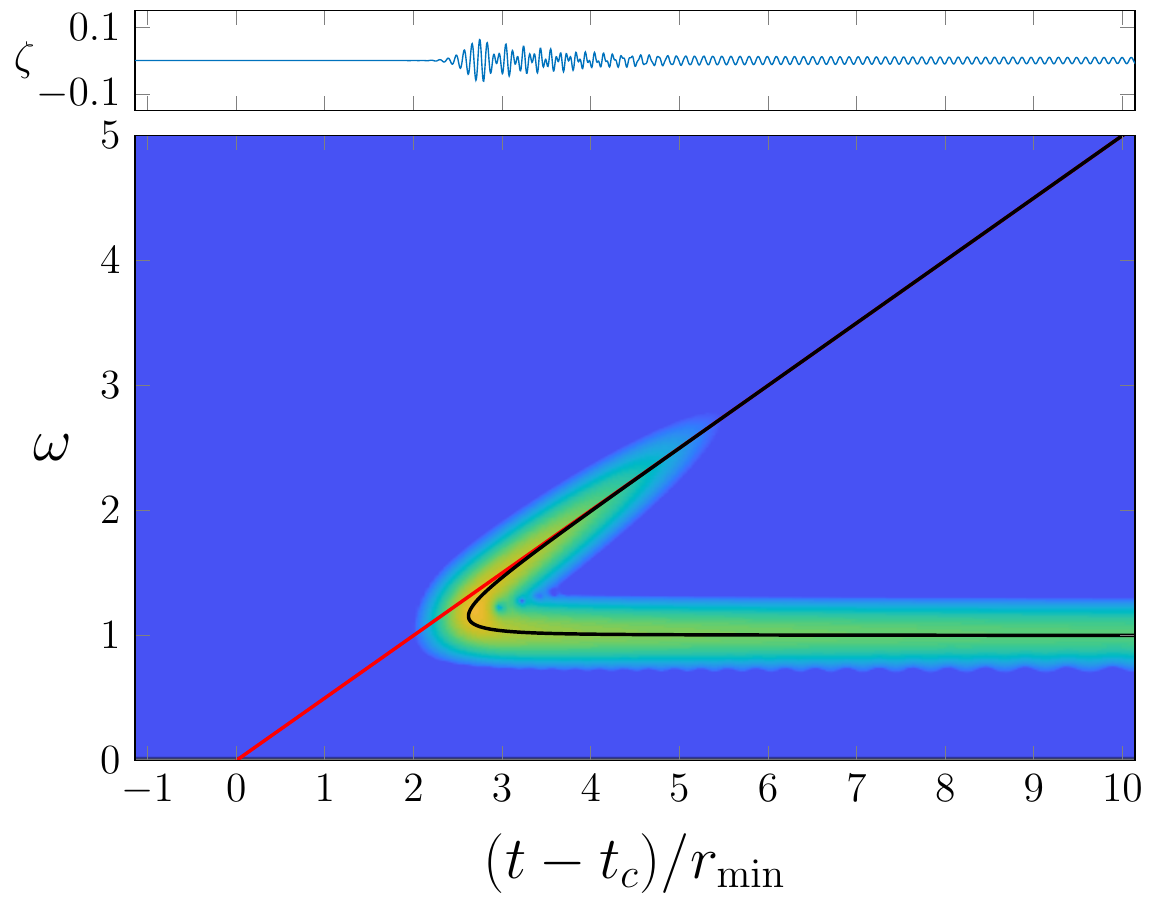}}&\\
	\end{tabular}
	\caption{(a,b) Shows the free-surface height, evaluated using (\ref{eq:surface}), for a pressure distribution starting at the point $(-1000,35)$ and moving with speed profile (\ref{eq:shipSpeedDecel}) for $\alpha=-0.01$ at two different times $t=t_c$ and $t_c+4r_\mathrm{min}$, the black dot is the location of the sensor. The ship comes to a stop at the point $(-50,35)$. (c--f) Spectrograms computed for acceleration $\alpha=-\infty,-0.25,-0.1,-0.01$ (note: $\alpha=-\infty$ refers to an impulsively stopped ship).The classical (black), and final-point (red) dispersion curves are overlaid.}
	\label{fig:LinSpecDecel}
\end{figure}

As a complement to Sec.~\ref{sec:accStraight} we will now examine spectrograms produced by a disturbance decelerating to a stop from a unit cruising velocity. The position and velocity vectors are given by (\ref{eq:straightPath}), while the ship's speed is
\begin{equation}
u(t)=
\begin{cases}
\mathrm{erf}\left(\frac{\sqrt{\pi}\alpha (t-t_s)}{2}\right) &t<t_s\\
0 & t\ge t_s
\end{cases},\label{eq:shipSpeedDecel}
\end{equation}
where $t_s$ is the time the ship stops and $\alpha<0$ is the final deceleration rate.  Wave patterns for this example are shown in Fig.~\ref{fig:LinSpecDecel}(a,b), drawn for the case $F=0.7$, $\alpha=-0.01$.  We see that this small amount of deceleration affects the wave pattern in an observable manner.  The divergent waves are cut off or diminished.  After the disturbance has stopped moving, the wave pattern is composed primarily of transverse waves.

Figure \ref{fig:LinSpecDecel} presents spectrograms for a decelerating disturbance with a final acceleration of $\alpha=-\infty,-0.25,-0.1,$ and $-0.01$ ($\alpha=-\infty$ refers to an impulsively stopped ship). All examples are for a disturbance that comes to a stop at $X(t_s)=(-50,35)$. As with Fig.~\ref{fig:LinSpec}, the classical dispersion curve is drawn in black; however, here the curve in red is the final-point dispersion curve.  As with the impulsively accelerated ship (Fig.~\ref{fig:LinSpec}(c)) the classical dispersion curve is a truncated version of the steady dispersion curve presented in Ref.~\cite{pethiyagoda17}.  The difference here is that it is the divergent branch that is truncated, not the transverse branch.  For finite values of the deceleration, the upper branch of the dispersion curves will still asymptotically approach the line $\omega=(t-t_c)/2r_\mathrm{min}$. This means that matching the asymptotes of the upper and lower branch to an experimental spectrogram will give the ship speed and its minimum distance to the sensor; however, it must be noted that this minimum distance may not be the same as the distance between the sensor and the sailing line (had the ship continued forward).

\subsection{Turning ship}
\begin{figure}
	\begin{tabular}{ccc}
		\subfloat[Surface $t=\pi R$]{\includegraphics[width=.45\linewidth]{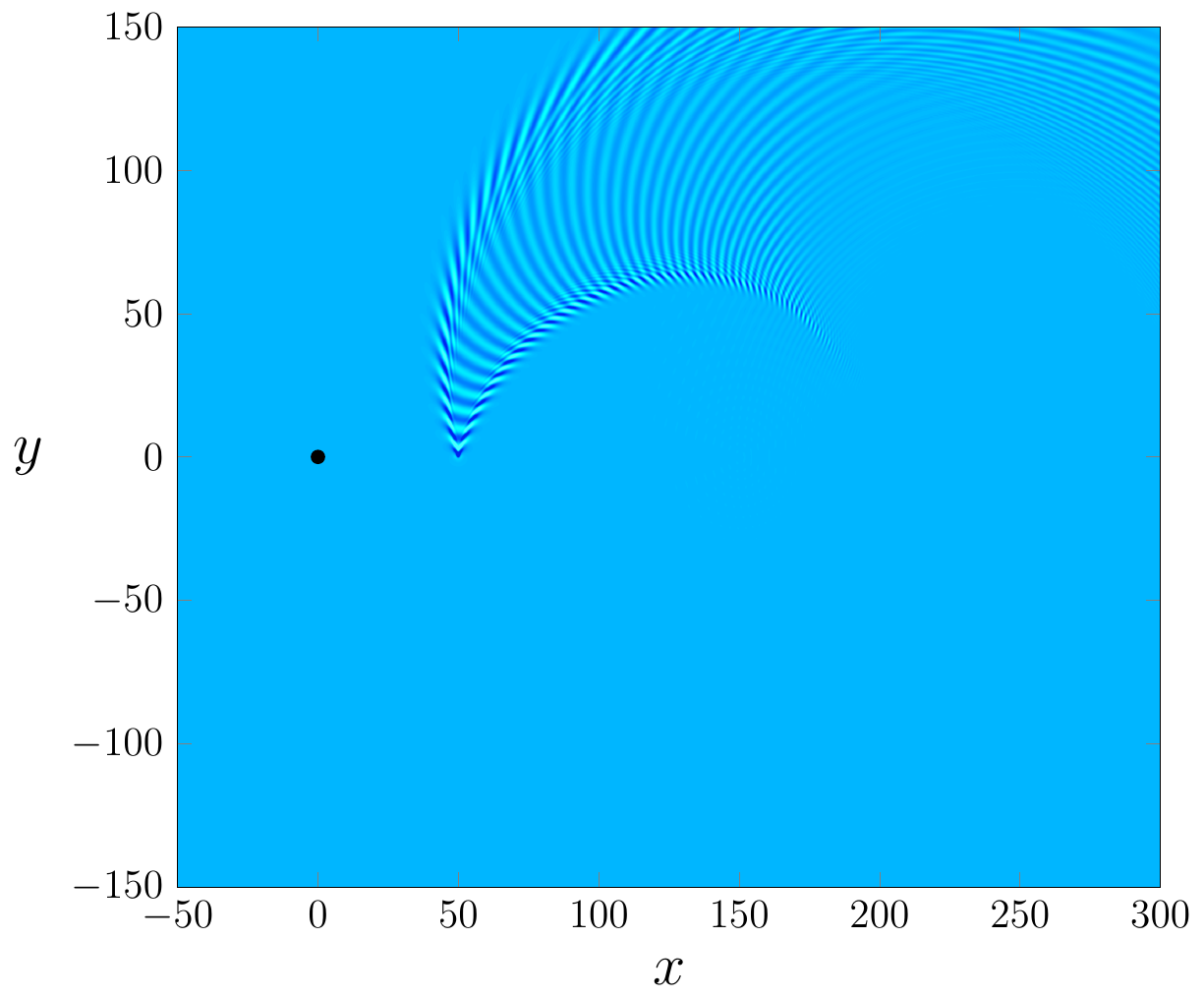}}&
		\subfloat[Surface $t=2\pi R$]{\includegraphics[width=.45\linewidth]{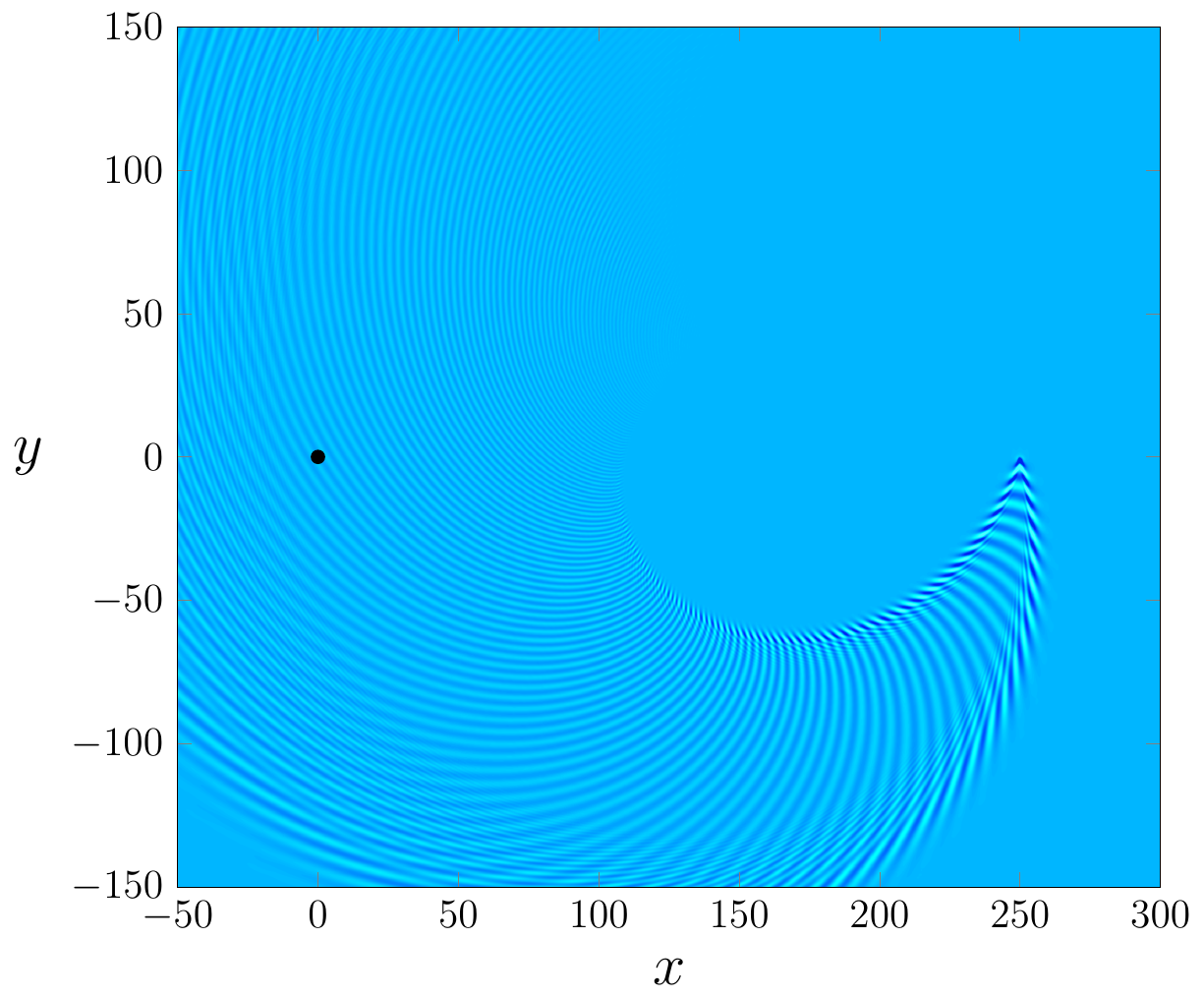}}&\\
		\subfloat[$R/x_c=1/3$]{\includegraphics[width=.45\linewidth]{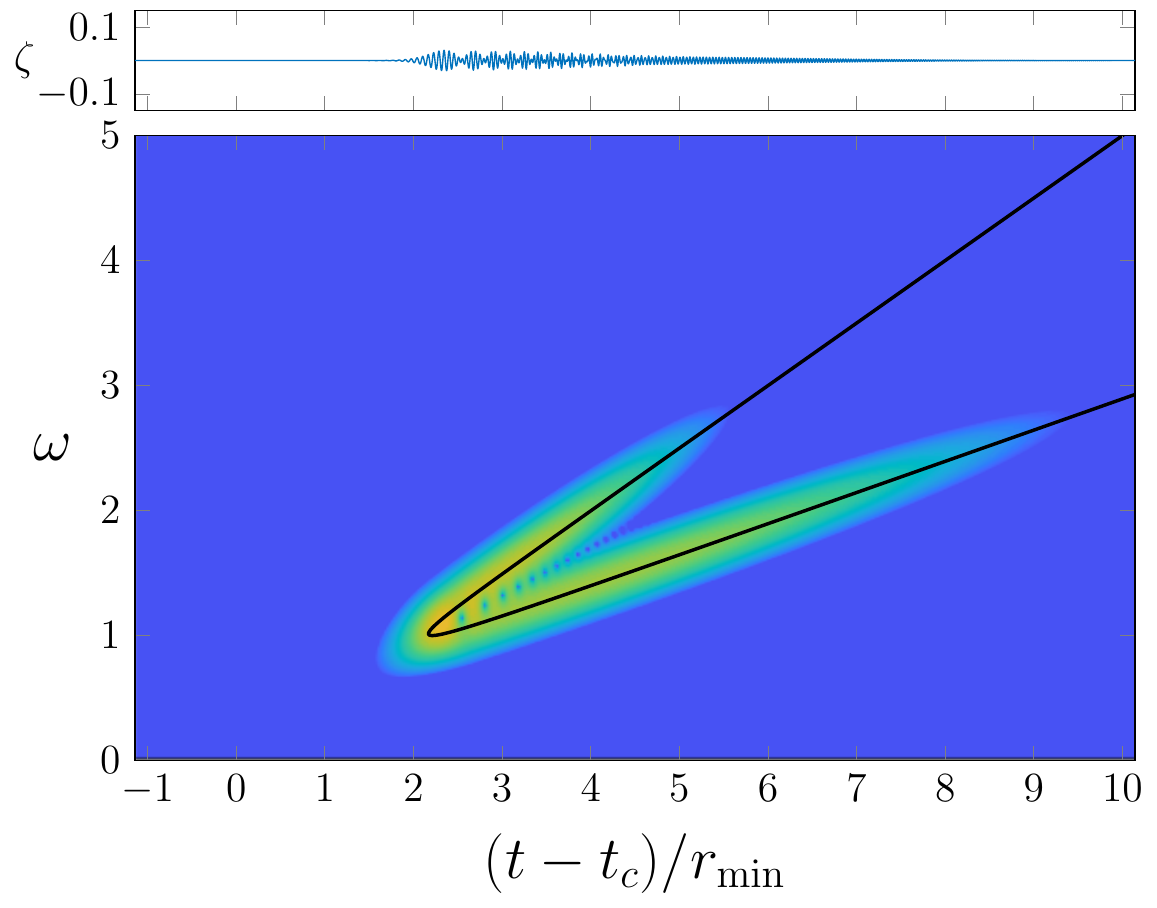}}&
		\subfloat[$R/x_c=2/3$]{\includegraphics[width=.45\linewidth]{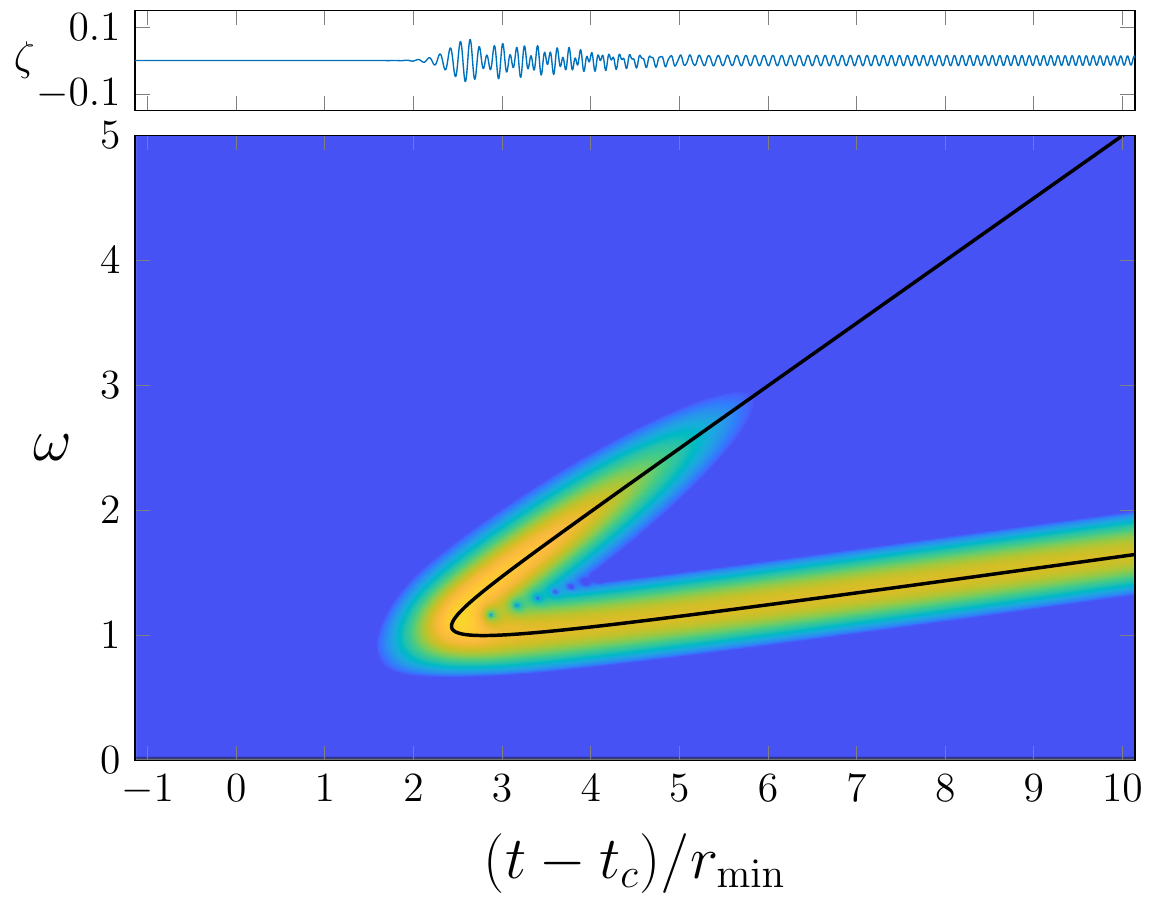}}&
		\hspace{-.0\linewidth}\multirow{4}{*}[0.17\linewidth]{\includegraphics[width=.06\linewidth]{colourbarLinear.pdf}}\\
		\subfloat[$R/x_c=4/3$]{\includegraphics[width=.45\linewidth]{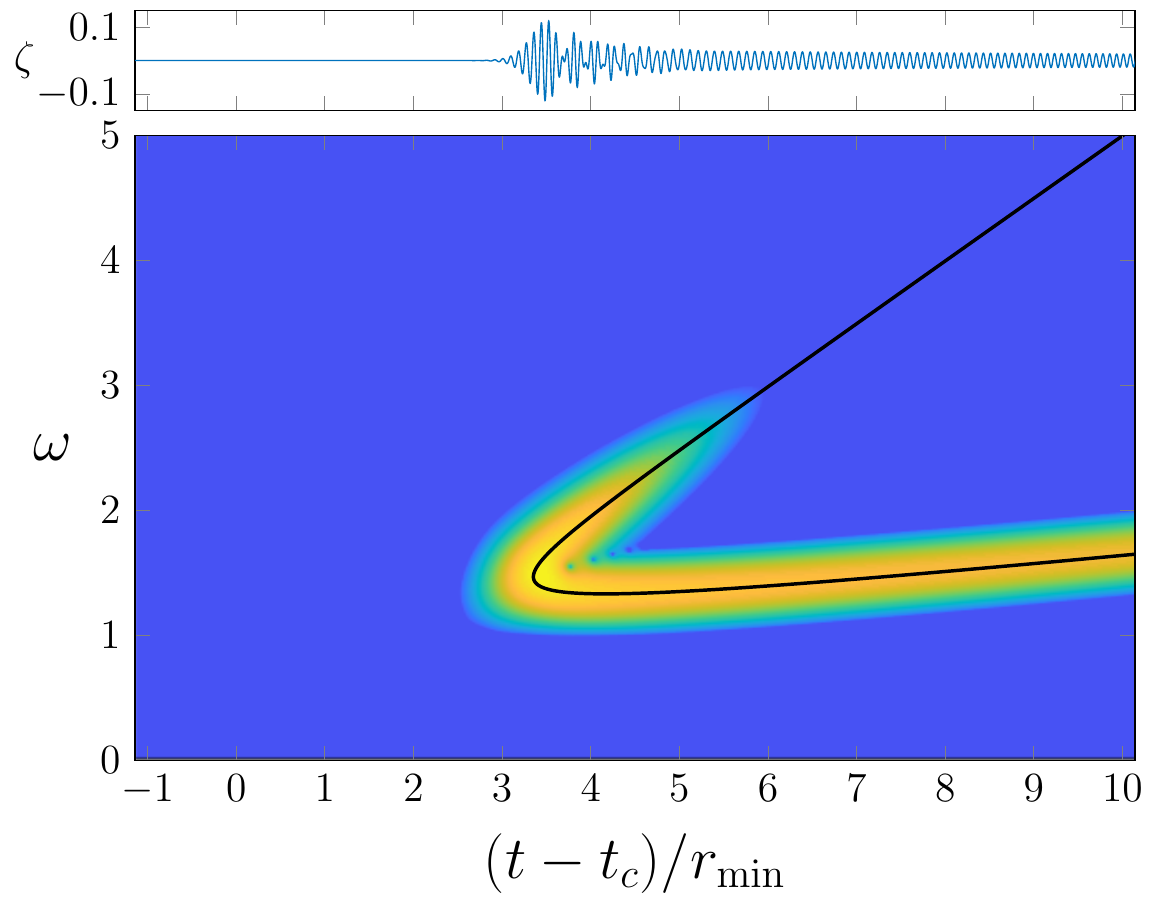}}&
		\subfloat[$R/x_c=5/3$]{\includegraphics[width=.45\linewidth]{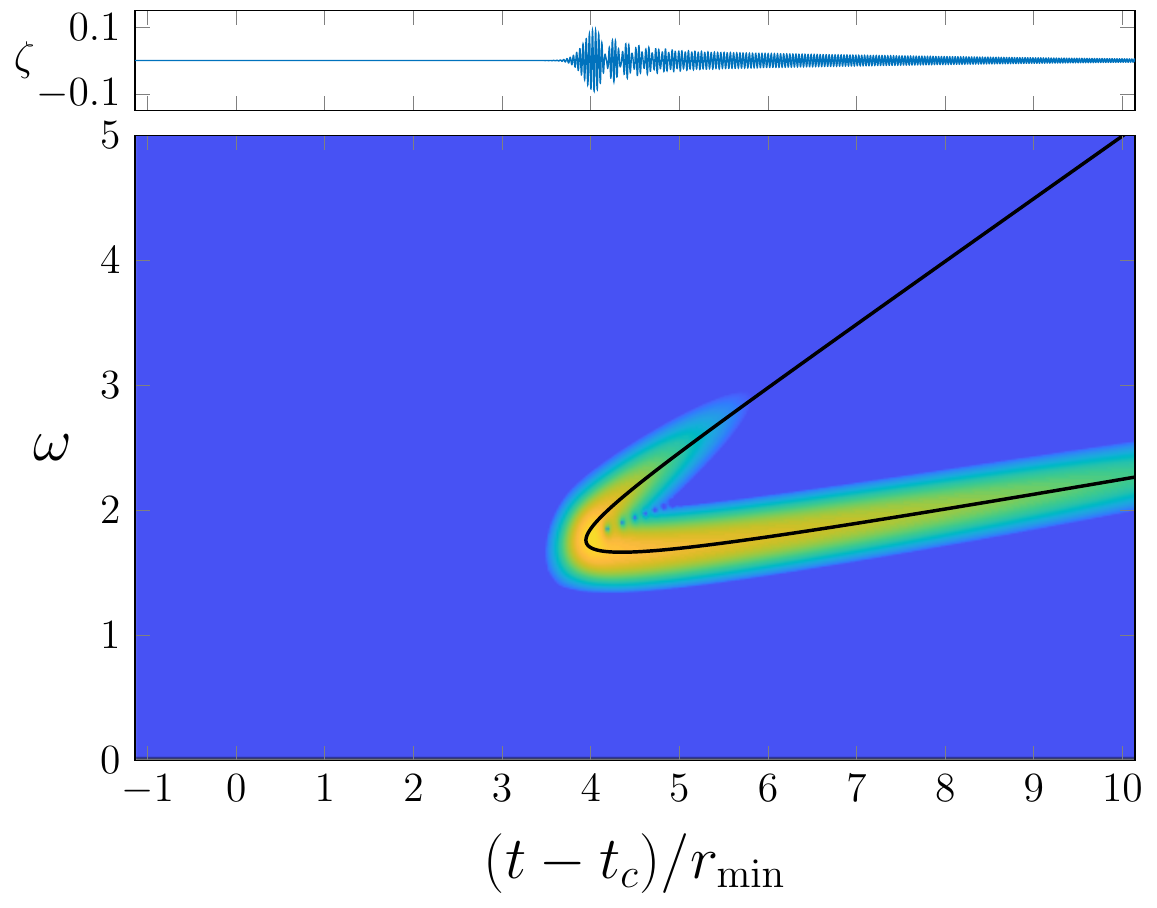}}&\\
	\end{tabular}
	\caption{(a,b) Shows the free-surface height evaluated using (\ref{eq:surface}) for a pressure distribution travelling in a circle with speed unity and turning radius $R=100$ at two different times $t=\pi R, 2\pi R$, the black dot is the location of the sensor. (c--f) Spectrograms computed for the same disturbance travelling in a circle with speed unity and circle centre $x_c=150$. The sensor positioned either (c,d) outside or (e,f) inside the turning circle. The classical dispersion curve is overlaid.}
	\label{fig:LinSpecTurn}
\end{figure}

For the next example, we will consider a ship moving with unit dimensionless angular speed in a circle of radius $R$ (the magnitude of the acceleration is $\alpha=1/R$). The ship's movement is given by
\begin{equation}
\textbf{X}(t)=
\begin{cases}
\left(x_c+R\cos\frac{t}{R},R\sin\frac{t}{R}\right) &0<t<2\pi R\\
\left(x_c+R,0\right)&\text{otherwise}
\end{cases},\qquad
\textbf{U}(t)=
\begin{cases}
\left(-\sin\frac{t}{R},\cos\frac{t}{R}\right) &0<t<2\pi R\\
\left(0,0\right)&\text{otherwise}
\end{cases},\label{eq:circlePath}
\end{equation}
where $x_c$ is the centre of the turning circle. This example was treated briefly in Ref.~\cite{Pethiyagoda2019}. As a representative solution, Fig.~\ref{fig:LinSpecTurn}(a,b) presents the wake pattern for a turning ship with $F=0.7$ and turning radius $R=100$ ($\alpha=0.01$). We observe the familiar wake pattern being distorted onto the circular path by compressing the divergent waves within the circle and stretching the divergent waves outside the circle.  Additionally, the transverse waves appear to fan out from the centre.

It can be shown that the dispersion curves for infinite-depth flow are identical for constant values of $R/x_c$ when plotted on our chosen scaled axis. We will present solutions for when the sensor is outside the turning circle $R/x_c<1$ and inside the turning circle $R/x_c>1$.

Figure \ref{fig:LinSpecTurn} presents spectrograms for a turning ship where the sensor is located (c,d)  outside and (e,f) inside of the turning circle, with the (black) classical dispersion curve overlaid. In all cases, as with the examples in Sec.~\ref{sec:accStraight}, the classical dispersion curve is comprised of two branches with the divergent branch approaching the line $\omega=(t-t_c)/2r_\mathrm{min}$ and the transverse branch approaching the final-time dispersion curve given by (\ref{eq:pulseDisp}) where $\tau_e$ is the time the ship is farthest away from the sensor (because this is when the ship's movement starts contributing to the dispersion curve). Additionally, for $R/x_c<1$, the classical dispersion curve has a minimum frequency of $\omega=1$ which occurs at the time when the ship is moving directly towards the sensor.

\subsection{Non-uniqueness of the dispersion curve}

As can be seen in Figs.~\ref{fig:LinSpec} and \ref{fig:LinSpecTurn} the classical dispersion curves for a ship accelerating in a straight line and a turning ship both have two rising frequency branches. Therefore, it seems possible for a ship travelling in a straight line to produce the same dispersion curve as a turning ship.  Indeed, we can determine the values $x_0$, $y_0$ and $u(t)$ in (\ref{eq:straightPath}) that will produce the same dispersion curve by equating the radial distance $r(t)=|\textbf{X}(t)|$ for $0<t<\pi R$ of a ship moving in a straight line (\ref{eq:straightPath}) and a turning ship (\ref{eq:circlePath}) to give
\begin{align}
x_0 = -2\sqrt{x_cR},\qquad
y_0 = x_c-R,\qquad
u(t) =
\begin{cases}
0,&t<0\\
\sqrt{\frac{x_c}{2R}}\frac{\sin\frac{t}{R}}{\sqrt{\cos\frac{t}{R}+1}},&0\leq t<\pi R\\
\sqrt{\frac{x_c}{R}},&t\geq\pi R
\end{cases}.\label{eq:sCircVel}
\end{align}

\begin{figure}
	\centering
	\subfloat[Path velocity]{\includegraphics[width=.5\linewidth]{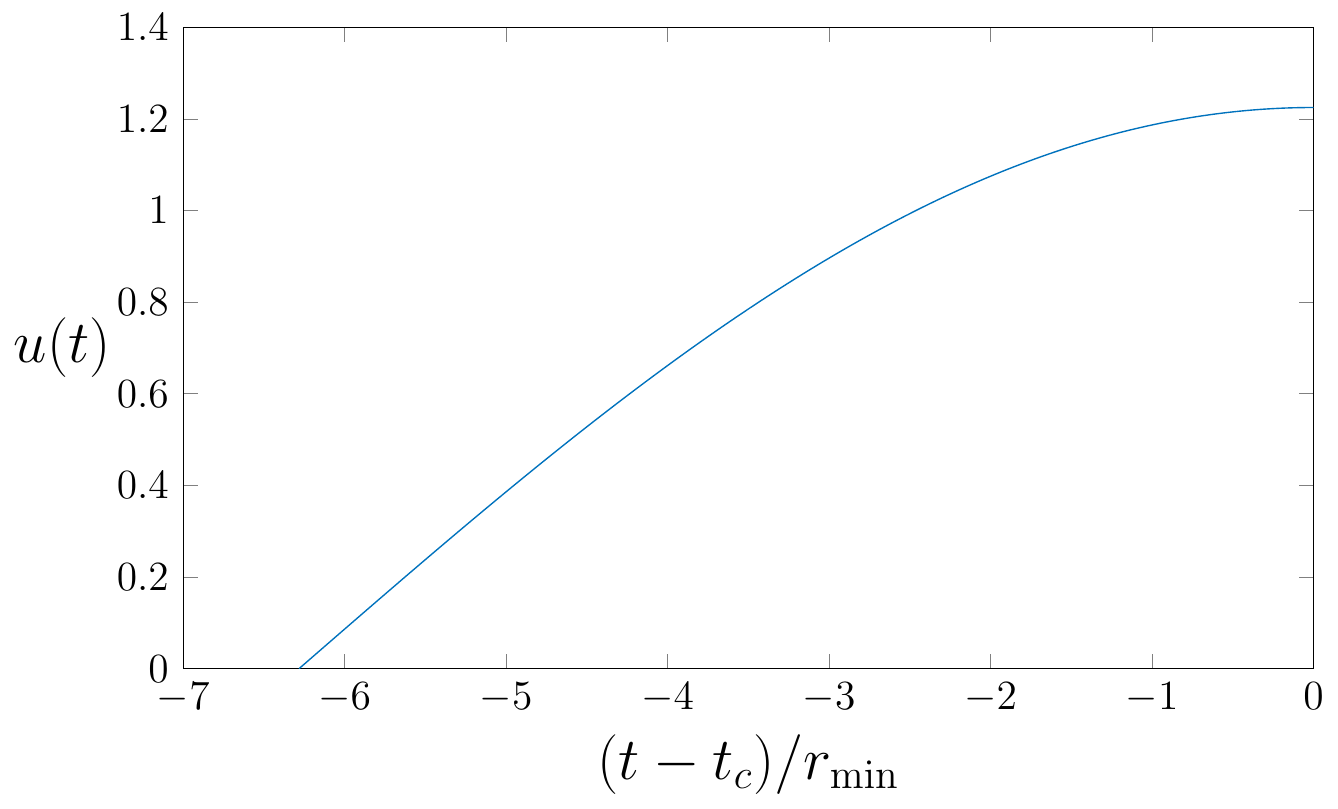}}\\
	\subfloat[Accerating ship free-surface]{\includegraphics[width=.53\linewidth]{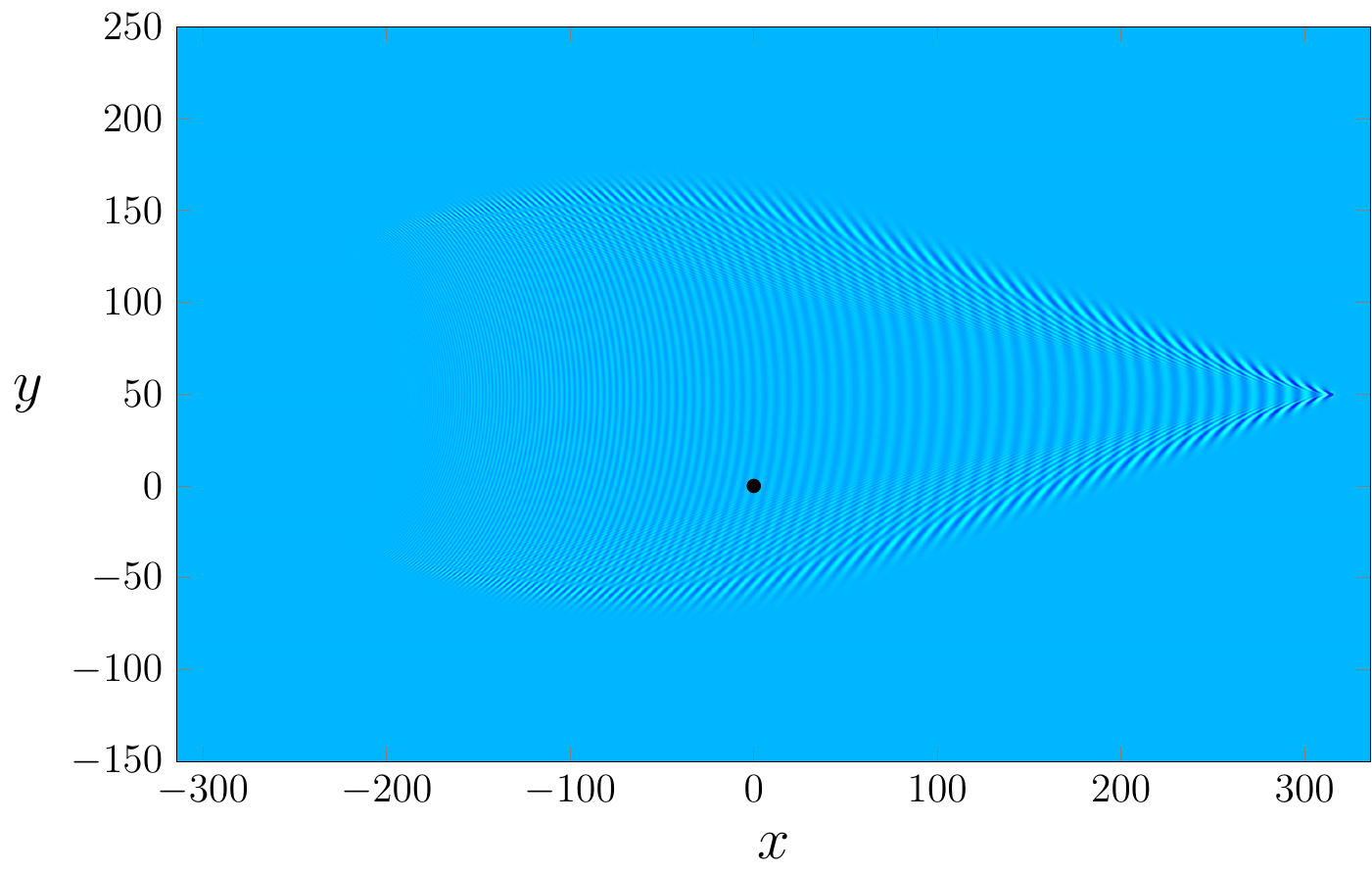}}\hspace{2ex}
	\subfloat[Turning ship free-surface]{\includegraphics[width=.4\linewidth]{Surf_circ_t2piR.pdf}}\\
	\subfloat[Accerating ship spectrogram]{\includegraphics[width=.45\linewidth]{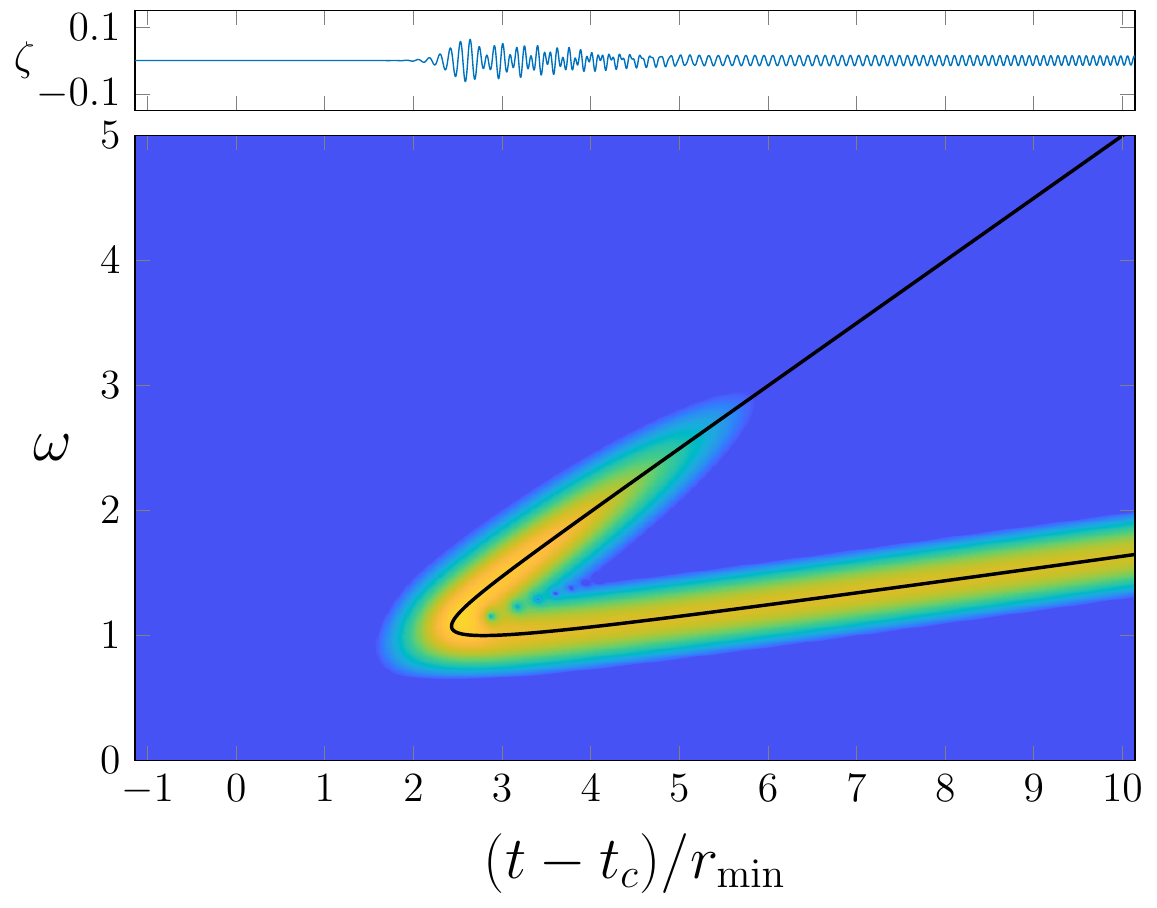}}
	\subfloat[Turning ship spectrogram]{\includegraphics[width=.45\linewidth]{SpecC_rad100.pdf} \raisebox{6ex}{\includegraphics[width=.0435\linewidth]{colourbarLinear.pdf}}}
	\caption{(a) A plot of the velocity profile for a ship accelerating in one direction along with (b) the free surface height. (c) The free-surface height for the turning ship with the same dispersion curve as (b) (Reproduction of Figure \ref{fig:LinSpecTurn}(b)). The black dot indicates the location of the sensor. The associated spectrograms are given for the (d) accelerating (e) turning ship (Reproduction of Figure \ref{fig:LinSpecTurn}(d)).}
	\label{fig:MatchingDisps}
\end{figure}

Figure \ref{fig:MatchingDisps} shows (a) the computed velocity profile for the accelerating ship with (b) its free-surface profile at $t=200\pi$ and (c) the free-surface profile for the associated turning ship for reference where $x_c=150$ and $R=100$. The signals for these two ships are presented with their spectrograms in Fig.~\ref{fig:MatchingDisps}(d) and (e). While the produced signals are not technically identical, both the signals and their associated spectrograms are visually indistinguishable.  This agreement is remarkable because the wave patterns themselves in Figure \ref{fig:MatchingDisps} (b) and (c) are clearly very different.  We have therefore demonstrated, via an example, how the issue of non-uniqueness could make it difficult to unpick properties of a wavemaker simply from a single spectrogram.

The similarity of the signals in Figure \ref{fig:MatchingDisps} (d) and (e) is possible in part due to the axisymmetric nature of the pressure distribution.  In reality, ships are not axisymmetric and therefore the additional information from the ship's hull shape could possibly be used to differentiate between spectrograms for different sailing paths. To briefly explore the effects of hull aspect ratio, we consider the non-axisymmetric pressure distribution $p(x,y)=\epsilon\,\exp(-\pi^2F^4(x^2+y^2/\beta^2))$, where $\beta$ is the width-length aspect ratio of the disturbance~\cite{benzaquen14,lo21,moisy14b}. The Fourier transform of the pressure distribution is $\tilde{p}(k,\psi)=\epsilon\beta\exp(-k^2(\cos^2\psi+\beta^2\sin^2\psi)/4\pi^2F^4)/(\pi F^4)$ and the surface profile is given by
\begin{align*}
\zeta(\textbf{x},t)&=\zeta_i(r(\textbf{x},0),\theta(\textbf{x},0),t)+\int_0^{t}\zeta_p(r(\textbf{x},\tau),\theta(\textbf{x},\tau),t-\tau)\,\mathrm{d}\tau,\\
\zeta_i(r,t)&=-\frac{1}{2\pi}\int_0^\infty \frac{k^2}{\Omega(k)^2}f(k)\cos\left(\Omega(k)t\right)\,\mathrm{d}k,\\
\zeta_p(r,t)&=-\frac{1}{2\pi}\int_0^\infty \frac{k^2}{\Omega(k)}f(k)\sin\left(\Omega(k)t\right)\,\mathrm{d}k,\\
f(k)&=\frac{\beta}{\pi F^4}\mathrm{e}^{-\frac{k^2(1+\beta^2)}{8\pi^2F^4}}\left[I_0\!\!\left(\frac{k^2(1-\beta^2)}{8\pi^2F^4}\right)J_{0}\!\left(kr\right) +2\sum_{n=1}^{\infty}I_n\!\!\left(\frac{k^2(1-\beta^2)}{8\pi^2F^4}\right)J_{2n}\!\left(kr\right)\cos({2n\theta})\right],
\end{align*}
where $I_n(x)$ is the modified Bessel function of the first kind of order $n$.

\begin{figure}
	\centering
	\subfloat[Accerating ship]{\includegraphics[width=.45\linewidth]{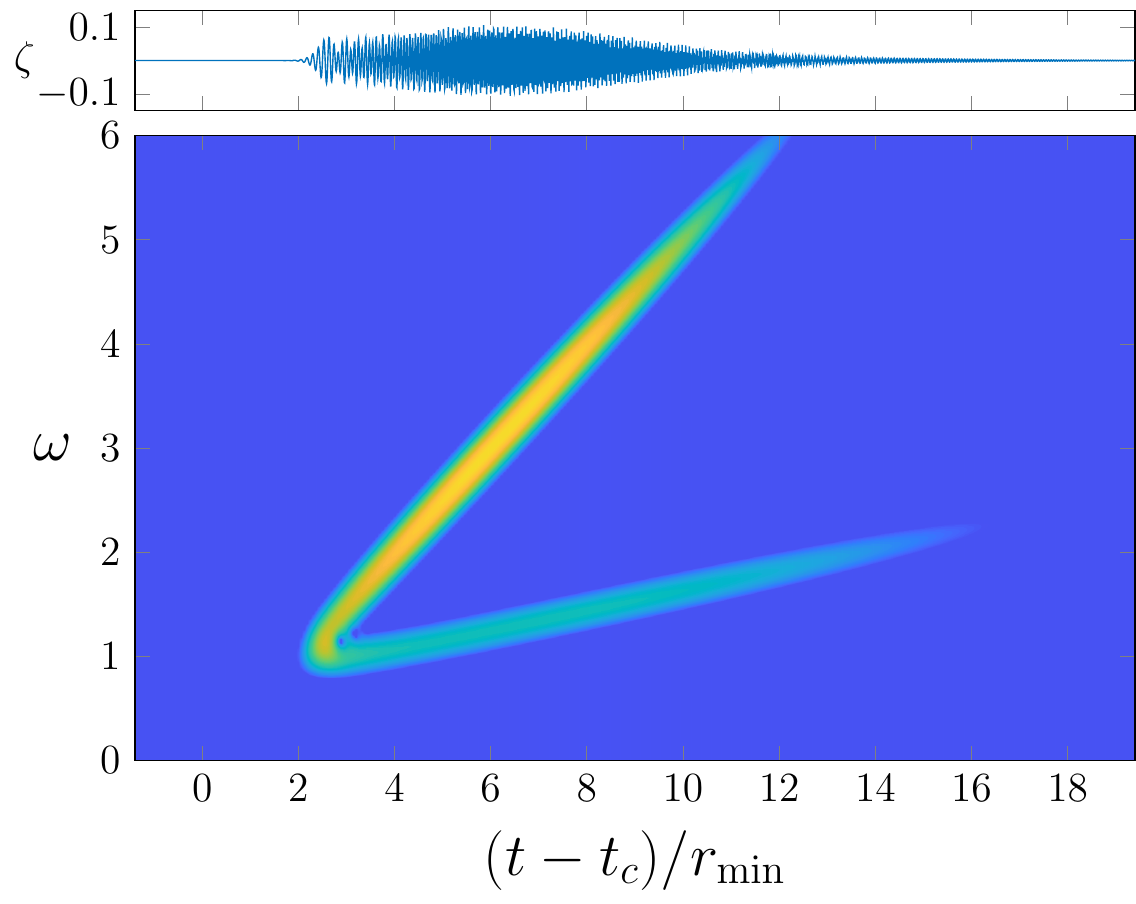}}\hspace{.5ex}
	\subfloat[Turning ship]{\includegraphics[width=.45\linewidth]{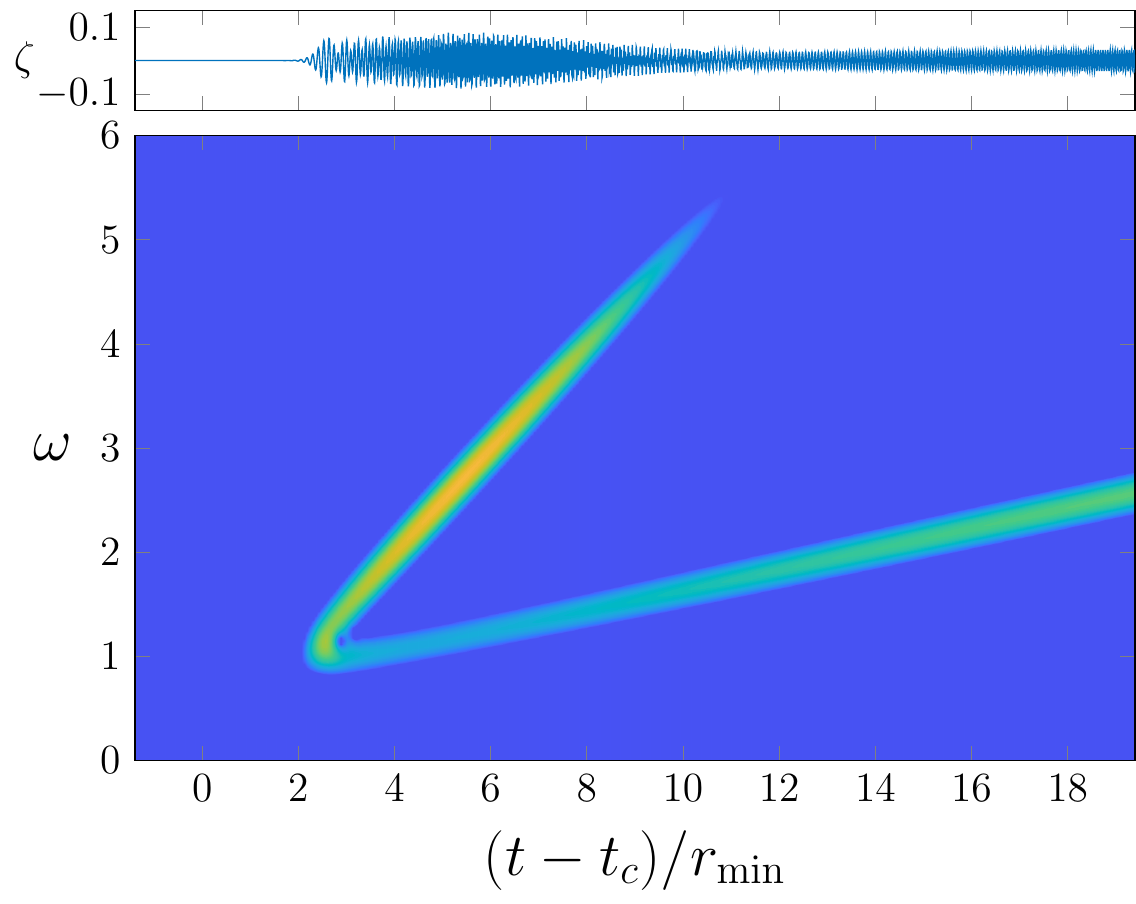} \raisebox{6ex}{\includegraphics[width=.0545\linewidth]{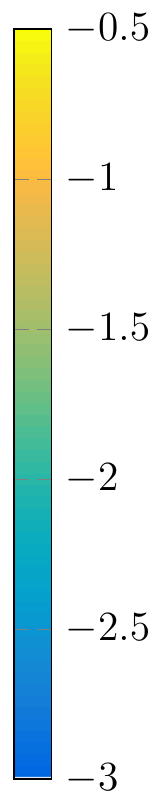}}}
	\caption{Spectrograms for a non-axisymmetric pressure distribution with aspect ratio $\beta=1/8$ (a) accelerating in a straight line and (b) turning in a circle.}
	\label{fig:nonaxiSpec}
\end{figure}

Figure \ref{fig:nonaxiSpec} presents spectrograms for the non-axisymmetric pressure distribution with aspect ratio $\beta=1/8$ for the same sailing paths shown in Fig.~\ref{fig:MatchingDisps}(d) and (e). The spectrogram colour intensities in Fig.~\ref{fig:nonaxiSpec} follow the same dispersion curve (not shown), as expected; however, the intensity along the dispersion curves differ between the sailing paths. For the ship accelerating in a straight line, the spectrogram has a maximum colour intensity along the upper branch and a monotonically decreasing intensity along the lower branch. The spectrogram for the ship moving  a circle also exhibits a maximum colour intensity along the upper branch; however, unlike the straight-path spectrogram, there is a local minimum of intensity along the lower branch. This local minimum corresponds the when the ship is travelling directly towards the sensor. Therefore, we see from this example that specific features in the spectrogram colour intensity, such as this local minimum, could be used to infer additional information about the sailing path of the ship. { Of course, on open water the presence of noise in the signal could obscure these distinguishing features, although wind waves can occupy higher frequency bands of the spectrogram than ship waves and are therefore relatively easy to identify \cite{didenkulova13}.}

 We note the non-uniqueness of the dispersion curve is to be expected from sampling the free-surface elevation at a single fixed location.  From (\ref{eq:sCircVel}), we see that the velocity of the accelerating ship only approaches unity if the associated turning ship sails directly into the sensor. This feature means that if a turning ship wake is misidentified as an accelerating ship, the cruising speed of the ship will be overestimated (underestimated) by the spectrogram analysis if the sensor is outside (inside) the turning circle.

{  As a final comment in this subsection, the example we highlight was constructed by equating the radial distance functions for the straight line and circular paths; however, any two paths that exhibit the same radial distance $r(t)$ will produce the same linear dispersion curve.} Therefore, the true nature of the ship's sailing path must be verified before performing any analysis, either by pre-existing knowledge of the environment (eg.\ a shipping channel where no turning is expected) or perhaps using information from multiple sensors { (e.g. triangulation)} or the distribution of colour intensity along the dispersion curve.

\section{Finite-depth effects and experimental results}\label{sec:finite}

In this section, we compare our theoretical predictions against experimental data collected in a model test basin.   As finite-depth effects play a role in these experiments, we will briefly present spectrograms and dispersion curves for accelerating disturbances where we take into account the effects of fluid depth.  This unsteady theory generalises that outlined in Ref.~\cite{pethiyagoda18b}, which was for steadily moving disturbances in a finite-depth channel.

\subsection{Theoretical results}

The surface elevation for a moving disturbance in a finite-depth channel is calculated using (\ref{eq:surface}) with
\begin{align*}
\zeta_p(r,t)&=-\frac{1}{2\pi}\int_0^\infty \frac{k^2\tanh(k/F_H^2)}{\Omega(k)}\tilde{p}(k)\sin\left(\Omega(k)t\right)J_0(kr)\,\mathrm{d}k,\\
\zeta_i(r,t)&=-\frac{1}{2\pi}\int_0^\infty \frac{k^2\tanh(k/F_H^2)}{\Omega(k)^2}\tilde{p}(k)\cos\left(\Omega(k)t\right)J_0(kr)\,\mathrm{d}k,
\end{align*}
where the dispersion function is $\Omega(k)=\sqrt{k\tanh(k/F_H^2)}$ (and, as mentioned earlier, $F_H=U/\sqrt{gH}$ is the depth-based Froude number).

\begin{figure}
		\subfloat[Straight path]{\includegraphics[height=.345\linewidth]{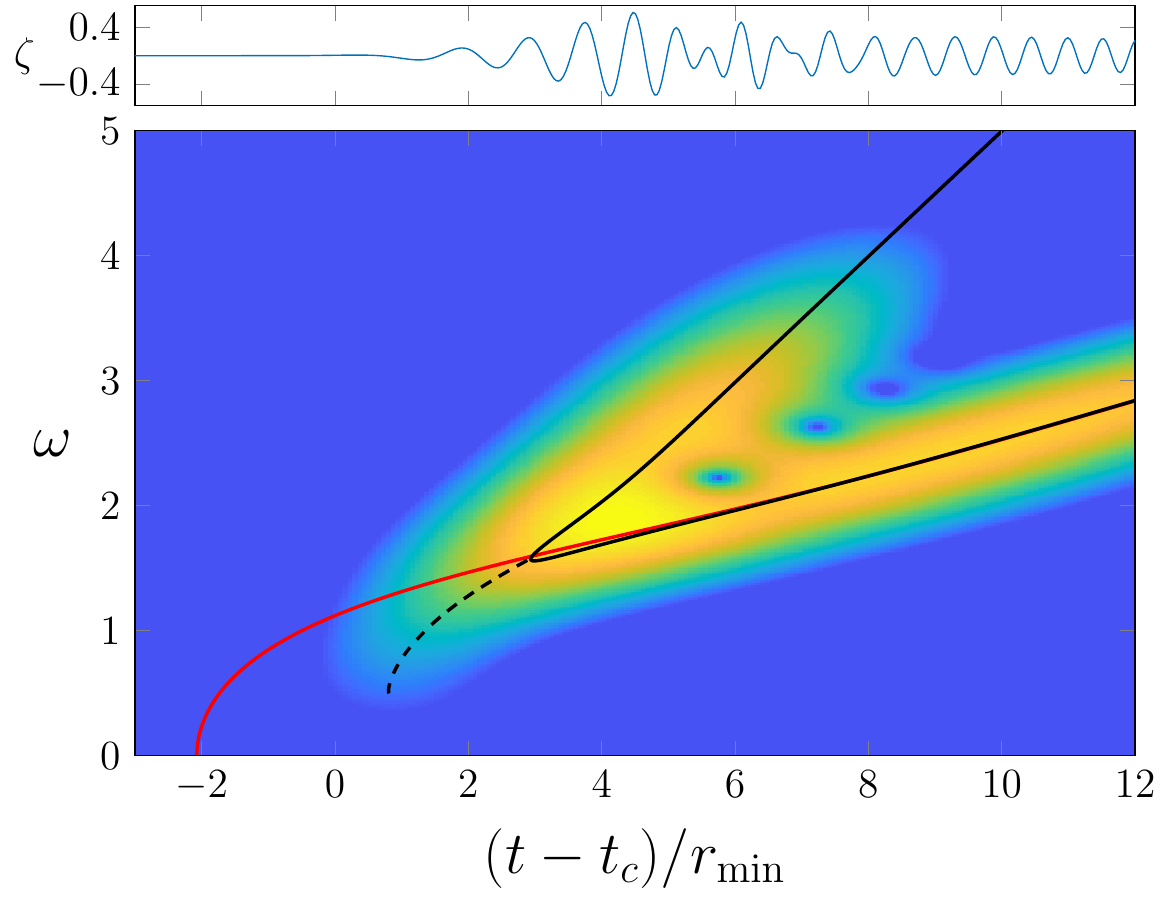}}\hspace{-1.5ex}
		\raisebox{2.6em}{\includegraphics[width=.052\linewidth]{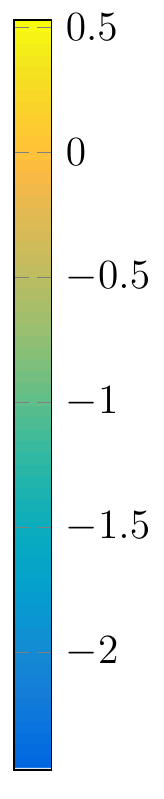}}
		\subfloat[Circular path]{\includegraphics[height=.345\linewidth]{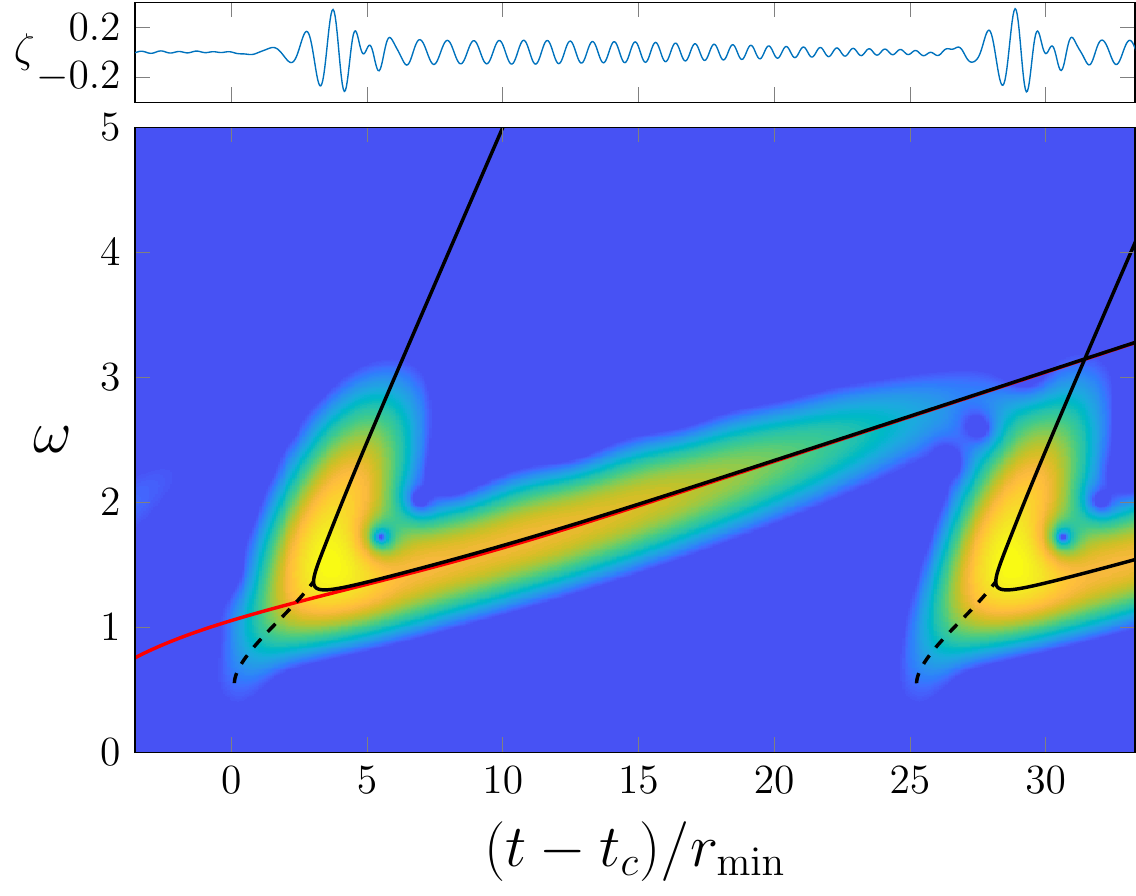}}
		\raisebox{2.5em}{\includegraphics[width=.053\linewidth]{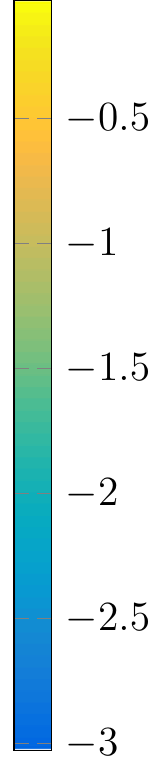}}
		\caption{Spectrograms for a pressure moving over a finite depth fluid (a) in a straight line starting at the point $(-13.08,4.36)$ with constant acceleration $\alpha=0.038$, and $F_H=1.23$, $F=1.14$, (b) turning in a circle with $x_c=12.49$, $R=16.65$, $F_H=0.89$ and $F=0.82$. The modified (black), and start-point (red) dispersion curves are overlaid.}\label{fig:LinFinite}
\end{figure}

We consider two example paths.  The first path is a ship moving in a straight line with constant acceleration (\ref{eq:straightPath}), where $u(t)=\alpha t$. Figure \ref{fig:LinFinite}(a) shows the spectrogram associated with this path, for the dimensionless acceleration $\alpha=0.038$, chosen to match the experimental data in Fig.~\ref{fig:ExpStraight}(a).  In this spectrogram, the classical linear dispersion curve (solid black curve), with its divergent and transverse branches, appears to do a good job of predicting the highest intensity colour regions.

A close inspection of Fig.~\ref{fig:LinFinite}(a) suggests there is a significant region of high colour intensity to the left of, and below, the fold in the classical dispersion curve.  A similar region of high colour intensity was identified by Torsvik \emph{et al.}~\cite{torsvik15a} in their experimental spectrograms taken from data collected in open water; these authors speculated that these low-frequency waves corresponded to precursor waves propagating ahead of the ship.  Whether or not there is a direct connection with real precursor waves, it is clear that these waves appear outside of the caustic and are not picked up by the classical dispersion curve, nor the start-point dispersion curve (solid red curve).

We attempt to explain the low-frequency waves in Fig.~\ref{fig:LinFinite}(a) by deriving a modified dispersion curve as follows.  Instead of solving (\ref{eq:cpPhys}) or (\ref{eq:cpTheta}) for $k$, which is not possible outside of the caustic, we set $k$ such that
\[
\min\limits_k\lvert kr^\prime(\tau)+\Omega(k)\rvert\quad\text{or}\quad\min\limits_k\left\lvert kU\cos(\psi)+\Omega(k)\right\rvert,
\]
respectively. That is, we move as close as possible to the solutions of the second equation in (\ref{eq:diffrelmid}). In terms of the method of stationary phase, instead of forcing the derivatives $\partial g_{1,2}/\partial\tau$ to vanish, we minimise them.  This is equivalent to generating an asymptotic expansion in the neighbourhood of the caustic, which would lead to an Airy-function representation of the surface \cite{chester57}.  This modified dispersion curve, which can be thought of as an extension of the classical dispersion curve, has an additional segment which we have drawn as a black dashed curve in Fig.~\ref{fig:LinFinite}(a).  This new component shows excellent agreement with the low-frequency region of the spectrogram.

The second path we consider is a ship turning in a circle (\ref{eq:circlePath}) without stopping, which gives rise to the spectrogram in Fig.~\ref{fig:LinFinite}(b).  Here we have chosen the parameter set $F_H=0.89$, $F=0.82$, $x_c=12.49$, and $R=16.65$, to match with our experimental results in Fig.~\ref{fig:ExpCirc}.  We see in Fig.~\ref{fig:LinFinite}(b) the classical dispersion curve (black curve) is an excellent predictor of the main location of high intensity colour in the spectrogram.  We observe a second copy of the classical dispersion curve as the ship turns a full circle and continues on.  Also drawn in Fig.~\ref{fig:LinFinite}(b) is the (black dashed) modified dispersion curve, which accounts for low-frequency waves outside of the caustic.  Note the (red) start-point dispersion curve is included to demonstrate that these low-frequency waves cannot be ascribed to transient effects from the initial impulse in speed.

\subsection{Experimental results}\label{sec:experiments}

Results from two model-scale experiments are presented here: a ship hull moving in a straight line with constant acceleration, and a ship following a circular path with constant angular velocity. The experiments were conducted at the Australian Maritime College in a shallow water basin using the AMC 00-01 model with waterline length of $1$m, beam of $0.227$m, and draught of $0.089$m (all measured when the model is at rest) \cite{macfarlane18}. The water depth in the basin is $0.3$m, which is small enough for finite-depth effects to be present.

For the constant-acceleration experiment, the hull was accelerated from rest at a rate of $0.375$ms$^{-2}$ over a distance of $12$m.  Two sensors were placed at a lateral distance of $2$m from the sailing line, $5$m (S1) and $11$m (S2) from the initial leading edge of the model. Sensor S1 observed the hull moving predominately at subcritical speeds, while sensor S2 observed both sub- and supercritical waves. The signals are nondimenionalised as outlined in Sec.~\ref{sec:specSteady}, where $U$ is given by the hull speed when it is closest to the sensor.

\begin{figure}
	\centering
	\subfloat[Sensor S1]{\includegraphics[width=.45\linewidth]{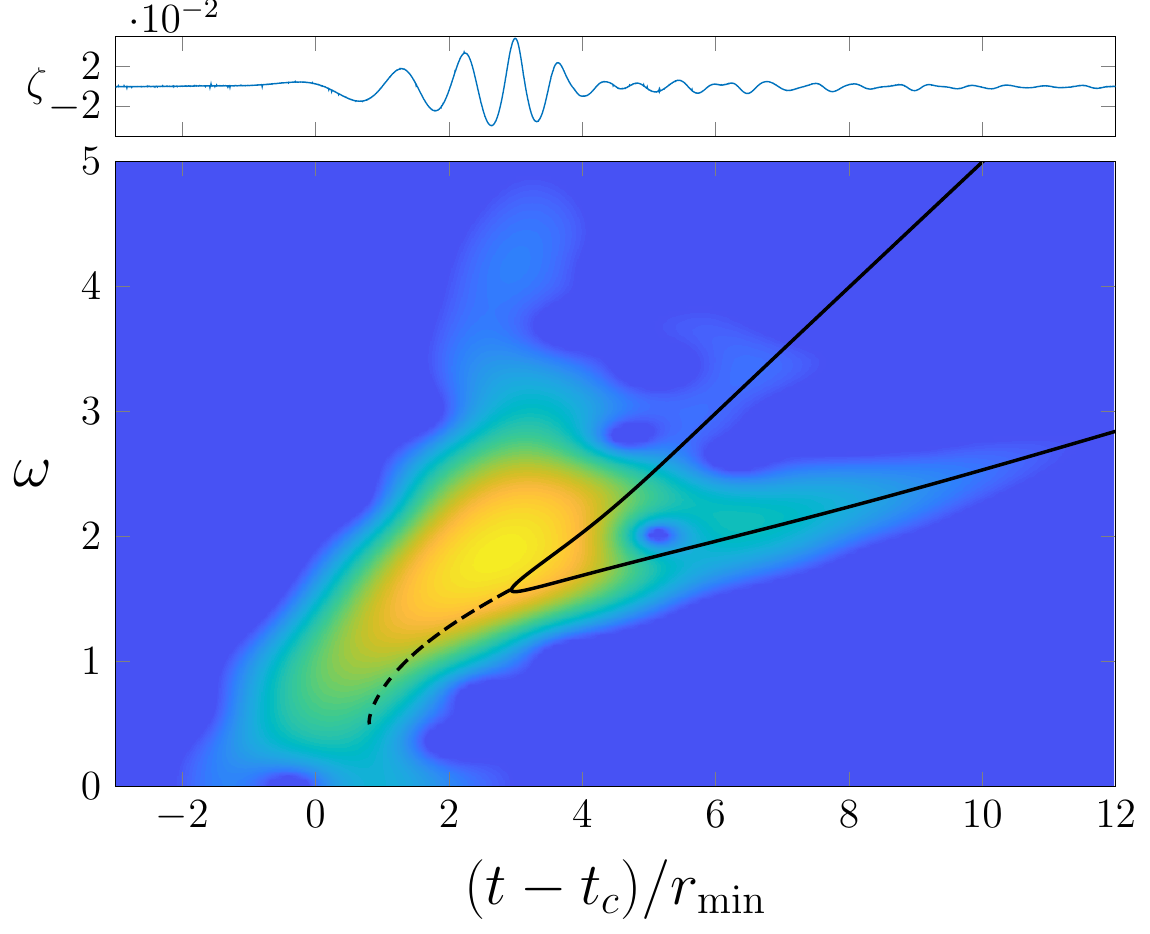}}\hspace{1ex}
	\subfloat[Sensor S2]{\includegraphics[width=.45\linewidth]{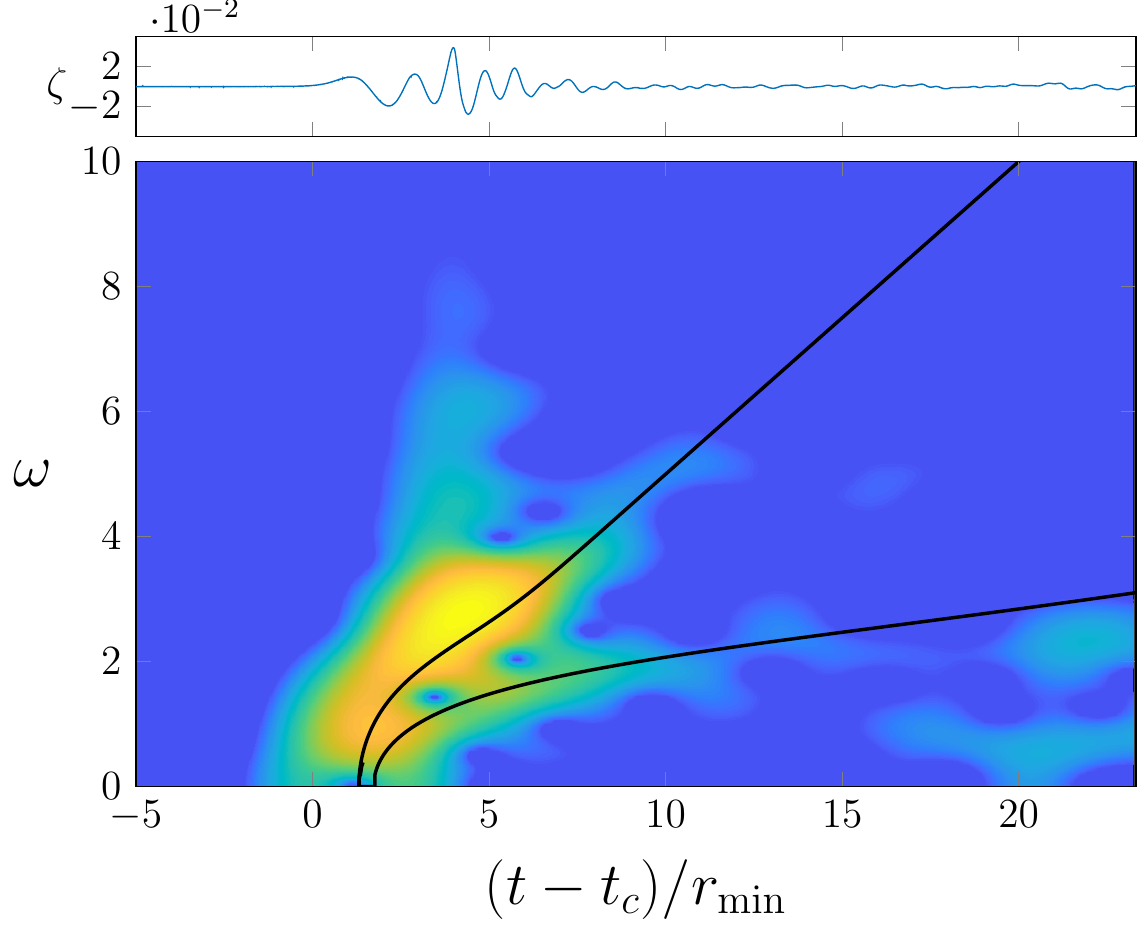}}\hspace{1ex}
	\raisebox{2.5em}{\includegraphics[width=.055\linewidth]{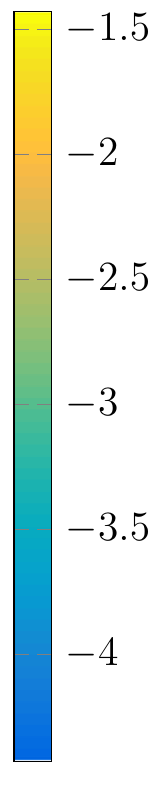}}
	\caption{Spectrograms computed from experimental measurements taken for a model undergoing constant acceleration of $0.375$ms$^{-2}$ by sensors (a) S1 and (b) S2 positioned $5$ and $11$ metres downstream from the initial location of the bow, respectively. The modified dispersion curve is overlaid.}
	\label{fig:ExpStraight}
\end{figure}

The spectrograms produced from the two sensors are presented in Fig.~\ref{fig:ExpStraight}. We can see that the (black) dispersion curve does a reasonable job of predicting the location of some of the colour intensity present in the spectrogram. However, we note that for the subcritical spectrogram (Fig.~\ref{fig:ExpStraight}(a)), to correctly match the dispersion curve we had to assume that the disturbance started moving six metres upstream of the sensor.  This adjustment is equivalent to assuming that the main source of the waves is at the stern of the hull. The subcritical spectrogram shows a significant high intensity colour region corresponding to low frequency waves to the left, and below, the classical dispersion curve.  As with our theoretical example in Fig.~\ref{fig:LinFinite}(a)), we have overlaid a (black dashed) modified dispersion curve; in this case, the curve does not do as good a job of predicting the location of these low-frequency waves outside of the caustic, although its general behaviour is captured.

Finally, for both of the experimental spectrograms in Fig.~\ref{fig:ExpStraight}, there is a medium colour intensity region corresponding to high frequency waves above the leading edge of the dispersion curves.  These waves are caused by nonlinearity, as discussed in \cite{pethiyagoda17,torsvik15a,pethiyagoda18b}, and so cannot be predicted by our linear theory.

\begin{figure}
	\centering
	\includegraphics[width=.45\linewidth]{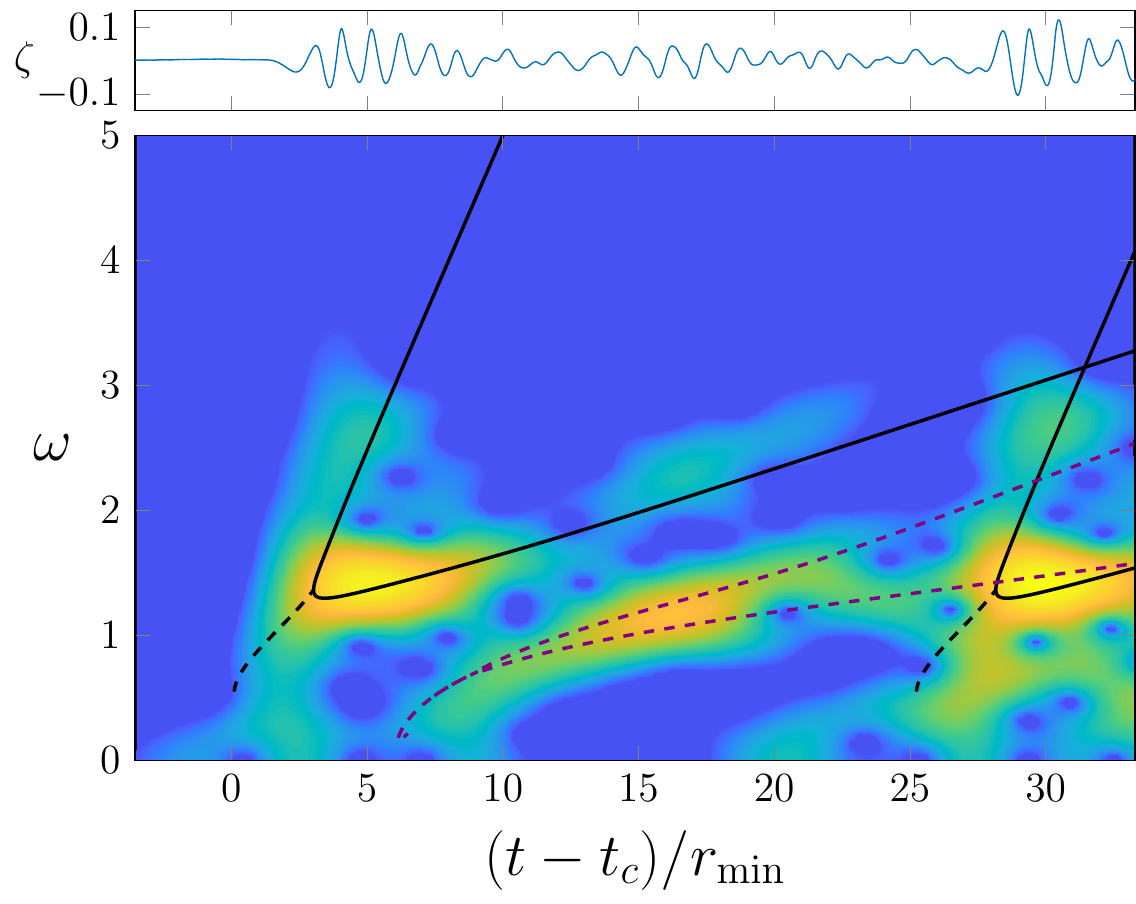}\hspace{1ex}
	\raisebox{2.5em}{\includegraphics[width=.055\linewidth]{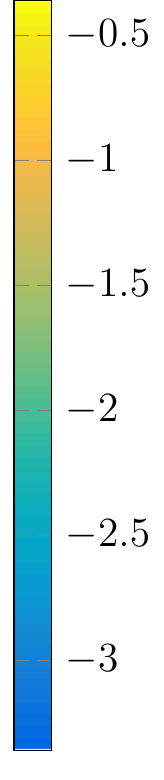}}
	\caption{A spectrogram computed from experimental measurements taken for a model moving in a circular path of radius $4$m with a constant speed of $1.535$ms$^{-1}$. The modified (black), and reflected (violet-dashed) dispersion curves are overlaid.}
	\label{fig:ExpCirc}
\end{figure}

The second model-scale experiment involved a hull moving along a circular path of radius $4$m, with a constant angular speed of $1.535$ms$^{-1}$ and a sensor placed $3$m from the centre of the circle. Figure \ref{fig:ExpCirc} shows the spectrogram for the measured wave elevation signal with the (black) classical dispersion curve again providing good agreement with the high colour intensity.  As with Fig.~\ref{fig:LinFinite}(b), more than one revolution was performed, resulting in a second classical dispersion curve with the same shape as the first being included.  We also note that, as with  Figs.~\ref{fig:LinFinite}(b) and \ref{fig:ExpStraight}(a), Fig.~\ref{fig:ExpCirc} also shows the low-frequency waves before the leading edge of the classic dispersion curve. However, the prediction from the (black dashed) modified dispersion curve is for higher frequency waves than what are observed.

The spectrogram in Fig.~\ref{fig:ExpCirc} is unfortunately complicated by a further region of colour intensity, which appears between the two classical dispersion curves.  This additional colour intensity region is due to wave reflection off the test basin walls.  Such reflection is reduced in the experimental set-up by using wave-damping ropes on the walls, however the small-amplitude reflecting waves are still detected by the sensors.  We have included in Fig.~\ref{fig:ExpCirc} a reflected dispersion curve as a violet dashed curve, generated by supposing there is a ``ghost'' ship travelling along a mirrored path.  This reflected curve shows very good agreement with the additional colour intensity in question.

\section{Discussion}\label{sec:discussion}

In this paper we explore the use of spectrograms to analyse surface elevation data collected from a single fixed sensor due to a ship accelerating along an arbitrary path, expanding upon our previous work for steadily moving ship.  This work is { an important step in developing the knowledge for} using spectrograms to decode these wave signals in a way that allows {  for the prediction of} features of { an accelerating} ship that causes the wake, { given that} not all shipping vessels are travelling at constant speed.  We use the method of stationary phase on a toy problem to determine the dispersion curve for these problems, which predicts the theoretical location of the highest colour intensity in a spectrogram. We further developed a simple geometric method for calculating the dispersion curve. For the examples presented in this paper (ie.\ a ship accelerating or decelerating in a straight line and a ship turning in a circle in infinite depth), the dispersion curve exhibited two rising frequency branches that { could potentially} be used to estimate physical features such as the minimum distance between the ship to the sensor and the ship's starting distance from the sensor. Unlike { the simpler problem} with steady ship waves, {  a proper analysis of the inverse problem must be conducted to determine the feasibility of reconstructing the speed of the ship from the dispersion curve.  We leave these issues related to the inverse problem for future research.}

We have highlighted a problem of non-uniqueness of spectrograms by constructing two very different paths (accelerating in a straight line and turning in a circle) that lead to identical linear dispersion curves. This non-uniqueness reinforces how difficult it is to back-calculate information about the cause of the disturbance based on this information alone.
Finally, we applied the dispersion curve calculated for finite-depth ship waves to linear ship waves and experimental data. While the dispersion curve showed good agreement with the measured spectrogram, there were additional regions of colour intensity that could be explained by properties such as nonlinearity and reflection, and a region that aligned with the previously identified precursor wave \cite{torsvik15a}. The comparison with experimental data shows that the linear dispersion curve for an accelerating ship can be applied to real world scenarios.

It is worth emphasising that the derivations in Sec.~\ref{sec:problemDef} involve a general wave function $\Omega(k)$ (more specifically, the wave signal derived in Sec.~\ref{sec:problemDef_I} holds for an infinitely deep fluid, while the theory in Sec.~\ref{sec:problemDef_II} and \ref{sec:problemDef_III} applies for any $\Omega(k)$).  As such, while we were focussed on infinite-depth and finite-depth surface gravity waves due to a moving ship, our analysis can easily be adapted for applications with different dispersion relations.  Typical examples that may be of interest include problems for which the effects of surface tension together with gravity are important \cite{chepelianskii08,moisy14a,raphael96}, models for weakly damped ship waves \cite{liang19}, hydroelastic waves such as those on floating ice sheet \cite{davys85,onoditbiot19}, internal or surface waves due to a stratified fluid \cite{li20,li2021}, and ship waves with constant vorticity \cite{ellingsen14,li16} (although further modifications would be required in the latter case due to $\Omega=\Omega(k,\psi)$).  Our results in Sec.~\ref{sec:problemDef} therefore have the potential for widespread usage.

{  The pressure distribution we use in our study creates a surface depression which changes shape with velocity and acceleration, qualitatively mimicking the effects of sinkage and trim.  The effect of using such a deformable hull in the model instead of a rigid hull is not clear, at least in terms of the spectrograms.  With this in mind,}
one of the obvious extensions of our work here and elsewhere \cite{pethiyagoda17,pethiyagoda18b} is to include a serious model for the shape of the ship hull and study how details of the hull shape appear in the time-frequency domain.  Preliminary results in that direction are reported in Ref.~\cite{Buttle2020}.  Questions remain about the role of wave interference with real ship hulls \cite{noblesse14,zhang15a}, as the combination of waves created at the bow and stern of a hull may lead to periodic patterns along the sliding frequency mode in a spectrogram heat map (such a phenomenon is hinted at in Ref.~\cite{pethiyagoda18b}).  There are a number of open problems regarding how nonlinearity manifests in the time-frequency domain, especially when it comes to steep surface gravity waves~\cite{wyatt88,parau02,pethiyagoda14a,pethiyagoda14b,buttle18} or weakly nonlinear waves propagating ahead of a ship travelling at near-critical speed~\cite{torsvik15a,soomere07,brown89}.  Finally, { as mentioned above,} the ultimate goal of this line of research is to devise algorithms to unpick experimental spectrograms so that, with only data collected from one or more fixed sensors, the key characteristics of the relevant vessel and its path may be predicted.  { This work is ongoing.}

\section{Acknowledgements}

The authors acknowledge the support of the Australian Research Council via the Discovery Project DP180103260 (SWM, TJM, GJM and RP) and the Linkage Project LP150100502 (GJM).  GJM acknowledges Keegan Graham-Parker for their assistance with the experimental work.  SWM would like to thank the Isaac Newton Institute for Mathematical Sciences, Cambridge, for support and hospitality during the programme Complex Analysis: Techniques, Applications and Computations where part of the work on this paper was undertaken.  This programme was supported by the EPSRC grant EP/R014604/1.  SWM is grateful for the generous support of the Simons Foundation who provided further financial support for his visit to the Isaac Newton Institute via a Simons Foundation Fellowship.  SWM acknowleges John Chapman from the University of Keele for insightful discussions on the method of stationary phase.  The authors thank the anonymous referees for their feedback.

\bibliographystyle{unsrt}
\bibliography{references}

\appendix
\section{Derivation of equation (\ref{eq:zetap})}\label{sec:derivingGovEq}
We begin by considering an inviscid, irrotational fluid bounded above by a free-surface, $z=\zeta(x,y,t)$, and below by a flat bottom of depth $h$ (that can be extended to infinity).  In such a dimensionless framework, $h=F_H^{-2}$, where $F_H$ is the depth-based Froude number.  The fluid is initially at rest and we apply a pressure disturbance, $p(x,y)$, to the surface as a single instance. We then linearise the system about the undisturbed free-surface, $z=0$. For the linearised system, the time-dependent velocity potential of the fluid, $\phi(x,y,z,t)$, satisfies Laplace's equation
\begin{equation}
\nabla^2\phi=0 \qquad\text{for } -h<z<0,
\end{equation}
and is subject to the linear kinematic and dynamic conditions,
\begin{align*}
\zeta_t=\phi_z,\qquad\phi_t+\mathcal{D}\{\zeta\}+\delta(t)p(x,y)=0\qquad\text{on }z=0,
\end{align*}
where $\mathcal{D}$ is a differential operator formed from a linear combination of constants and partial derivatives (of any order) with respect to $x$ and $y$, and $\delta(t)$ is the Dirac delta function. For example, $\mathcal{D}=1-\gamma/(\partial_{xx}+\partial_{yy})$ for gravity-capillary waves, $\mathcal{D}=1$ for gravity waves and $\mathcal{D}=-\gamma/(\partial_{xx}+\partial_{yy})$ for capillary waves, where $\gamma$ is a dimensionless surface tension. We enforce no flow through the bottom boundary so that
\begin{align*}
\frac{\partial\phi}{\partial z}=0\qquad\text{on }z=-h.
\end{align*}
The initial conditions are
\[
\phi=\phi_t=\zeta=\zeta_t=0\qquad\text{at }t=0.
\]
Applying the two-dimensional Fourier transform
\[
\tilde{g}(k,\psi)=\int_0^{2\pi}\int_0^\infty g(x,y)\mathrm{e}^{-irk\cos(\theta-\psi)}r\,\mathrm{d}r\,\mathrm{d}\theta,
\]
where $r=\sqrt{x^2+y^2}$ and $\theta=\tan^{-1}(y/x)$ to the governing equations gives,
\begin{align}
-k^2\tilde{\phi}+\tilde{\phi}_{zz}&=0 &&\text{for } -h<z<0,\label{eq:laplace}\\
\tilde{\zeta}_t&=\tilde{\phi}_z&&\text{on }z=0,\label{eq:kin}\\
\tilde{\phi}_t+\tilde{\mathcal{D}}(k,\psi)\tilde{\zeta}+\delta(t)\tilde{p}(k,\psi)&=0&&\text{on }z=0,\label{eq:transDyn}\\
\frac{\partial\tilde{\phi}}{\partial z}&=0 &&\text{on }z=-h,\label{eq:bottom}
\end{align}
where $\tilde{\mathcal{D}}$ is a function defined so that $\tilde{\mathcal{D}}\tilde{\zeta}$ is the Fourier transform of $\mathcal{D}\{\zeta\}$. Solving (\ref{eq:laplace}) and applying the bottom boundary condition (\ref{eq:bottom}) gives
\[
\tilde{\phi}=A(k,\psi,t)\cosh{kz}+A(k,\psi,t)\tanh{kh}\sinh{kz}.
\]
Returning to the kinematic condition (\ref{eq:kin}), we have
\begin{equation}
\tilde{\zeta}_t=A(k,\psi,t)k\tanh{kh}=k\tilde{\phi}\tanh{kh}\qquad\text{on }z=0.\label{eq:tphiKin}
\end{equation}
Substituting (\ref{eq:tphiKin}) into (\ref{eq:transDyn}), we rearrange to get a second order ODE for $\tilde{\zeta}$
\begin{equation}
\tilde{\zeta}_{tt}+k\tanh{kh}\,\tilde{\mathcal{D}}\tilde{\zeta}=-\delta(t)k\tanh{kh}\,\tilde{p}(k,\psi),\label{eq:waveFullZeta}
\end{equation}
subject to the initial conditions
\begin{align}
\tilde{\zeta}(k,\psi,0)&=0,\\
\tilde{\zeta}_t(k,\psi,0)&=0.\label{eq:init}
\end{align}
Equations (\ref{eq:waveFullZeta})-(\ref{eq:init}) are equivalent to $\zeta$ satisfying the wave equation subject to an initial velocity given by $-k\tanh{kh}\,\tilde{p}(k,\psi)$ in phase space.  As such we let $\Omega(k,\psi)=k\tanh{kh}\,\tilde{\mathcal{D}}(k,\psi)$ be the dispersion function. Additionally, because we only consider isotropic free-surface effects in this paper, we can remove the dependence on $\psi$ from the dispersion function.

{ Thus, we can write the governing equations as
\begin{align}
\tilde{\zeta}_{tt}+\Omega(k)^2\tilde{\zeta} = -\delta(t)k\tanh{kh}\tilde{p}(k,\psi),\label{eq:zetaODE}\\
\tilde{\zeta}(k,\psi,0)=0, \quad
\tilde{\zeta}_t(k,\psi,0)=0.\label{eq:zetapinit}
\end{align}
Solving (\ref{eq:zetaODE})-(\ref{eq:zetapinit}) and inverting the Fourier transform gives the wave profile in polar coordinates,
\begin{equation}
\zeta(r,\theta,t)=-\frac{1}{4\pi^2}\int_0^\infty\int_{-\pi}^\pi \frac{k^2\tanh{kh}}{\Omega(k)}\tilde{p}(k,\psi)\sin\left(\Omega(k)t\right)\mathrm{e}^{\mathrm{i}kr\cos(\theta-\psi)}\,\mathrm{d}\psi\,\mathrm{d}k.\label{eq:zetap_ap}
\end{equation}
We recover the equation (\ref{eq:zetap}) by taking the depth out to infinity, $h\rightarrow\infty$.
}
\end{document}